\documentclass[aps,prl,twocolumn,floatfix,footinbib,showpacs,superscriptaddress]{revtex4-1}
\usepackage{graphicx}
\usepackage{amsfonts,amsmath,amssymb}
\usepackage{amsthm}
\usepackage{dsfont,bm}
\usepackage{color}
\usepackage{soul} 
\usepackage{amsbsy}
\usepackage[colorlinks=true,linkcolor=blue,pagecolor=blue,filecolor=blue,menucolor=blue,urlcolor=blue,citecolor=blue,anchorcolor=blue]{hyperref}

\usepackage{mathbbol}

\usepackage{sidecap}

\newcommand{\bsigma}{{\boldsymbol \sigma}}

\newcommand{\ac}{{\alpha_\text{c}}}
\newcommand{\bc}{{\beta_\text{c}}}

\begin{document}

\title{Cubic spin-orbit coupling and anomalous Josephson effect in planar junctions} 

\author{Mohammad Alidoust}
\affiliation{Department of Physics, Norwegian University of Science and Technology, N-7491 Trondheim, Norway}
\author{Chenghao Shen}
\affiliation{University at Buffalo, State University of New York, Buffalo, NY 14260-1500, USA}
\author{Igor \v{Z}uti\'c} 
\affiliation{University at Buffalo, State University of New York, Buffalo, NY 14260-1500, USA}
\date{\today}

\begin{abstract}
Spin-orbit coupling in two-dimensional systems is usually characterized by Rashba and Dresselhaus spin-orbit coupling (SOC) linear in the wave vector. However, there is a growing class of materials which instead support dominant SOC cubic in the wave vector (cSOC), while their superconducting properties remain unexplored. By focusing on Josephson junctions in Zeeman field with  superconductors separated by a normal cSOC region, we reveal a strongly anharmonic current-phase relation and complex spin structure. An experimental cSOC tunability enables both tunable anomalous phase shift and supercurrent, which flows even at the zero-phase difference in the junction. A fingerprint of cSOC in Josephson junctions is the $f$-wave spin-triplet superconducting correlations, important for superconducting spintronics and supporting Majorana bound states.  

\end{abstract}
\maketitle

Spin-orbit coupling (SOC) and its symmetry breaking provide versatile opportunities for materials design, brining relativistic phenomena to the fore of the condensed matter 
physics~\cite{Kane2005:PRL,Konig2007:S,Wan2011:PRB,Burkov2011:PRL,Armitage2018:RMP,Zutic2019:MT}. 
While for decades SOC was primarily studied to elucidate and manipulate normal-state 
properties, including applications in spintronics and quantum computing~\cite{Bychkov1984:PZETF,DasSarma2001:SSC,Winkler:2003,Zutic2004:RMP,Hanson2007:RMP,Fabian2007:APS,Xiao2010:RMP,Sinova2015:RMP,Schliemann2017:RMP}, 
there is a growing interest to examine its role on superconductivity~\cite{Gorkov2001:PRL,Samokhin2005:PRL,Reynoso2008:PRL,Buzdin2008:PRL,Eschrig2015:RPP,Smidman2017:RPP}. 

Through the coexistence of SOC and Zeeman field, a conventional spin-singlet superconductivity can acquire spin-dependent long-range proximity 
effects~\cite{Eschrig2015:RPP, Martinez2020:PRA,Jeon2020:PRX,Gonzalez-Ruano2020:PRB}
as well as support topological superconductivity and host Majorana bound states, a building block for fault-tolerant 
quantum computing~\cite{Lutchyn2010:PRL,Oreg2010:PRL,Aasen2016:PRX}. In both cases, Josephson junctions (JJs) provide a desirable platform to acquire spin-triplet superconductivity 
through proximity effects~\cite{Keizer2006:N,Robinson2010:S,Khaire2010:PRL,Banerjee2014:NC,Gingrich2016:NP,Linder2015:NP,Rokhinson2012:NP,Fornieri2019:N,Ren2019:N,Desjardins2019:NM,Mayer2019:P}. In contrast, even seemingly well-established intrinsic spin-triplet superconductivity in 
Sr$_2$RuO$_4$~\cite{Mackenzie2003:RMP} is 
now increasingly debated~\cite{Pustogow2019:N,Sharma2020:PNAS}.

Extensive normal-state studies of SOC in zinc-blende heterostructures  usually distinguishing the resulting spin-orbit fields due to broken bulk inversion symmetry, 
Dresselhaus SOC, and surface inversion asymmetry, Rashba SOC, and focus on their dominant linear dependence in the wave vector, 
${\bf k}$~\cite{Zutic2004:RMP,Schliemann2017:RMP}. In this linear regime, with a matching strengths of these 
SOCs it is possible to strongly suppress the spin relaxation~\cite{Schliemann2003:PRL} and realize a persistent spin helix (PSH)~\cite{Bernevig2006:PRL,Koralek2009:N} 
with a controllable spin precession over long distances~\cite{Dettwiler2017:PRX,Walser2012:NP,Iizasa2020:PRB}.  

\begin{figure}[b]
\centering
\includegraphics[width=11.5cm]{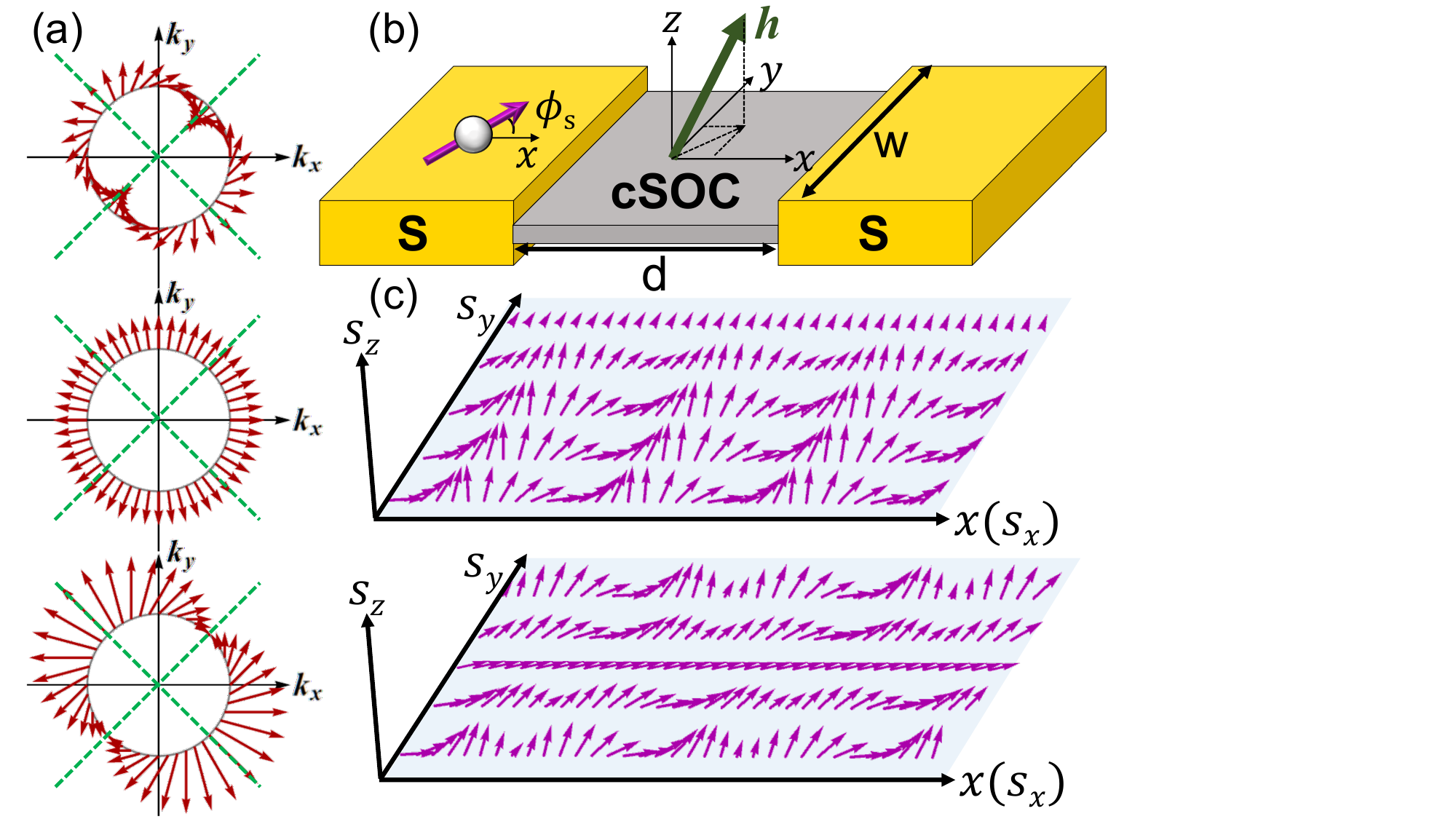}
\caption{(a) Spin-orbit fields in k-space for Rashba cubic spin-orbit coupling (cSOC) $(\alpha_c = -1$), Dresselhaus cSOC 
 ($\beta_c =-1$, middle), and both ($\alpha_c = \beta_c =-1$, bottom).
(b) Schematic of the Josephson junction. The middle region hosts cSOC and an effective Zeeman field, {\bf h},  
between  the two $s$-wave superconductors (S).
(c) Spin textures in the cSOC region resulting from the normal-incident electrons with in-plane spin orientations [see Fig.~\ref{fig1}(b)] when S is at normal-state, 
the upper (lower) panel  $\alpha_c = 1$, $\beta_c =0$ ($\alpha_c =\beta_c =1$). The in-plane spin orientations of the incident electrons $\phi_s$ are from 0 (bottom row) to 
$\pi/2$ (top row).}
\label{fig1}
\end{figure}

While typically  {\bf k}-cubic SOC contributions  (cSOC) in heterostructures are neglected or considered just detrimental perturbations, for example, limiting the stability of
 PSH~\cite{Dettwiler2017:PRX,Walser2012:NP,Iizasa2020:PRB}, a more complex picture is emerging. Materials advances suggest that such cSOC, shown in Fig.~\ref{fig1}(a), not only has to be included, but may also dominate the 
normal-state properties~\cite{Winkler2002:PRB,Krich2007:PRL,Altmann2006:PRL,Yoshizumi2016:APL,Kammermeier2016:PRL,Nakamura2012:PRL,Moriya2014:PRL,Cottier2020:PRB,Liu2018:PRL,Brosco2016:PRL}.
However, the role of cSOC in  superconducting heterostructures is unexplored. It is unclear if cSOC is detrimental or desirable to key phenomena such as Josephson effect, 
spin-triplet superconductivity, or Majorana bounds states. 

To address this situation and motivate further cSOC studies of superconducting properties, 
we consider JJs depicted in Fig.~\ref{fig1}(b), where $s$-wave superconductors 
(S) are separated by a normal region with cSOC which is consistent with the two-dimensional (2D) electron or hole gas, confined along the z-axis~\cite{Winkler2002:PRB,Nakamura2012:PRL}.   
We find that the interplay between Zeeman field and cSOC results in an anomalous Josephson effect with a 
spontaneous supercurrent. While the commonly-expected current-phase relation (CPR) is $I(\varphi)=I_c\sin(\varphi+\varphi_0)$~\cite{Buzdin2008:PRL,Strambini2020:NN}, 
where $I_c$ is the JJ critical current and $\varphi_0$ the anomalous phase ($\varphi_0 \neq 0, \pi$),
we reveal that CPR can be strongly anharmonic and host Majorana bound states. 
Instead of the $p$-wave superconducting correlations for  linear SOC, their $f$-wave symmetry is the fingerprint of cSOC.

To study cSOC, we consider an effective Hamiltonian 
\begin{equation}
H =\frac{1}{2}\int d\mathbf{p}~ \hat{\psi}^{\dag}(\mathbf{p}) H(\mathbf{p})\hat{\psi}(\mathbf{p}),
\end{equation}
where $H(\mathbf{p})  =  \mathbf{p}^2/2m^* +\bsigma\cdot \mathbf{h}+H_\text{cSOC}(\mathbf{p})$, 
with momentum, $\mathbf{p}=(p_x,p_y,0)$ [see Fig.~\ref{fig1}(b)], effective mass, $m^*$, Pauli matrices, ${\bsigma}$, effective Zeeman field, ${\bf h}$,
realized from an externally applied magnetic field or through magnetic proximity effect~\cite{Zutic2019:MT,Takiguchi2019:NP},
and the cSOC term~\cite{Winkler2002:PRB,Krich2007:PRL,Nakamura2012:PRL,Moriya2014:PRL}
\begin{equation}
H_\text{cSOC}(\mathbf{p}) = \frac{i\ac}{2\hbar^3}(p_-^3\sigma_+ - p_+^3\sigma_-)-\frac{\bc}{2\hbar^3}(p_-^2p_+\sigma_+  +  p_+^2p_-\sigma_-),
\end{equation}
expressed using cSOC strengths $\alpha_c$ and $\beta_c$, for  Rashba and Dresselhaus terms, 
where $p_\pm = p_x \pm i p_y$, and $\sigma_\pm = \sigma_x \pm i \sigma_y$. The field operator in spin space is given by $\hat{\psi}(\mathbf{p})=[\psi_{\uparrow}(\mathbf{p}), \psi_{\downarrow}(\mathbf{p})]^{\mathrm{T}}$, with $\uparrow, \,\downarrow$  spin projections.

To describe  S regions  
in Fig.~\ref{fig1}(b), we use an $s$-wave BCS model with a two-electron amplitude in spin-Nambu space $\Delta \langle \psi_\uparrow^\dag \psi_\downarrow^\dag \rangle + \text{H.c.}$, given by the effective Hamiltonian
in particle-hole space 
\begin{equation}\label{Hamil_sc}
{\cal H}(\mathbf{p}) = \left( \begin{array}{cc}
 H(\mathbf{p}) -\mu \hat{1}& \hat{\Delta} \\
 \hat{\Delta}^\dag & -H^\dagger(-\mathbf{p}) +\mu \hat{1}
\end{array}\right),
\end{equation}
where $\mu$  is the chemical potential and $\hat{\Delta} $ is a $2\times2$ gap matrix in spin space. 
The field operators in the rotated particle-hole and spin basis are $\hat{\psi}=(\psi_{\uparrow}, \psi_{\downarrow}, \psi_{\downarrow}^{\dag}, -\psi_{\uparrow}^{\dag})^{\mathrm{T}}$.

To calculate the charge current, we use its quantum definition where no charge sink or source is present. Therefore, the time variation of charge density vanishes, $\partial_t\rho_\text{c}\equiv 0=\lim\limits_{\mathbf{r}\rightarrow \mathbf{r}'}\sum\limits_{\sigma\tau\sigma'\tau'}[ \psi^\dag_{\sigma\tau}(\mathbf{r}'){\cal H}_{\sigma\tau\sigma'\tau'}(\mathbf{r})\psi_{\sigma'\tau'}(\mathbf{r})-\psi^\dag_{\sigma\tau}(\mathbf{r}'){\cal H}_{\sigma\tau\sigma'\tau'}^\dag(\mathbf{r}')\psi_{\sigma'\tau'}(\mathbf{r})]$. 
${\cal H}_{\sigma\tau\sigma'\tau'}$ is the component form of 
${\cal H}$, with spin (particle-hole) label $\sigma$ ($\tau$), and 
and $\mathbf{ r}\equiv(x,y,0)$. From the current conservation,  
the charge current density is,  
$\mathbf{ J} =\int \hspace{-.1cm} d\mathbf{r}\{\hat{\psi}^\dagger(\mathbf{r}) \overrightarrow{{\cal H}}(\mathbf{r})\hat{\psi}(\mathbf{r})-
\hat{\psi}^\dagger(\mathbf{r}) \overleftarrow{{\cal H}}(\mathbf{r})\hat{\psi}(\mathbf{r}) \},$ where ${\cal H}(\mathbf{r})$ is obtained by substituting $\mathbf{ p}\equiv -i \hbar (\partial_x,\partial_y,0)$. The arrow directions indicate the specific wavefunctions that the ${\cal H}$ 
operates on. 
By an exact diagonalization of ${\cal H}$, we obtain spinor wavefunctions $\hat{\psi}^{l,r,m}(\textbf{p})$ within the left (\text{$x<0$}) and right (\text{$x>d$})  S region 
and  the middle normal region 
(\text{$\smash 0<x<d$}) in Fig.~\ref{fig1}(b). The wavefunctions and generalized velocity operators $v_{x}^{l,r,m}$ are 
continuous at the junctions, i.e., $\hat{\psi}^l$=$\hat{\psi}^m|_{x=0}$, $\hat{\psi}^m$=$\hat{\psi}^r|_{x=d}$, $v_{x}^{l}\hat{\psi}^l$=$v_{x}^{m}\hat{\psi}^r|_{x=0}$, and $v_{x}^{m}\hat{\psi}^m$=$v_{x}^{r}\hat{\psi}^r|_{x=d}$. The spinor wavefunctions are given in the Supplemental Material~\cite{sm}. 

\begin{figure}[b]
\centering
\includegraphics[width=0.485\textwidth]{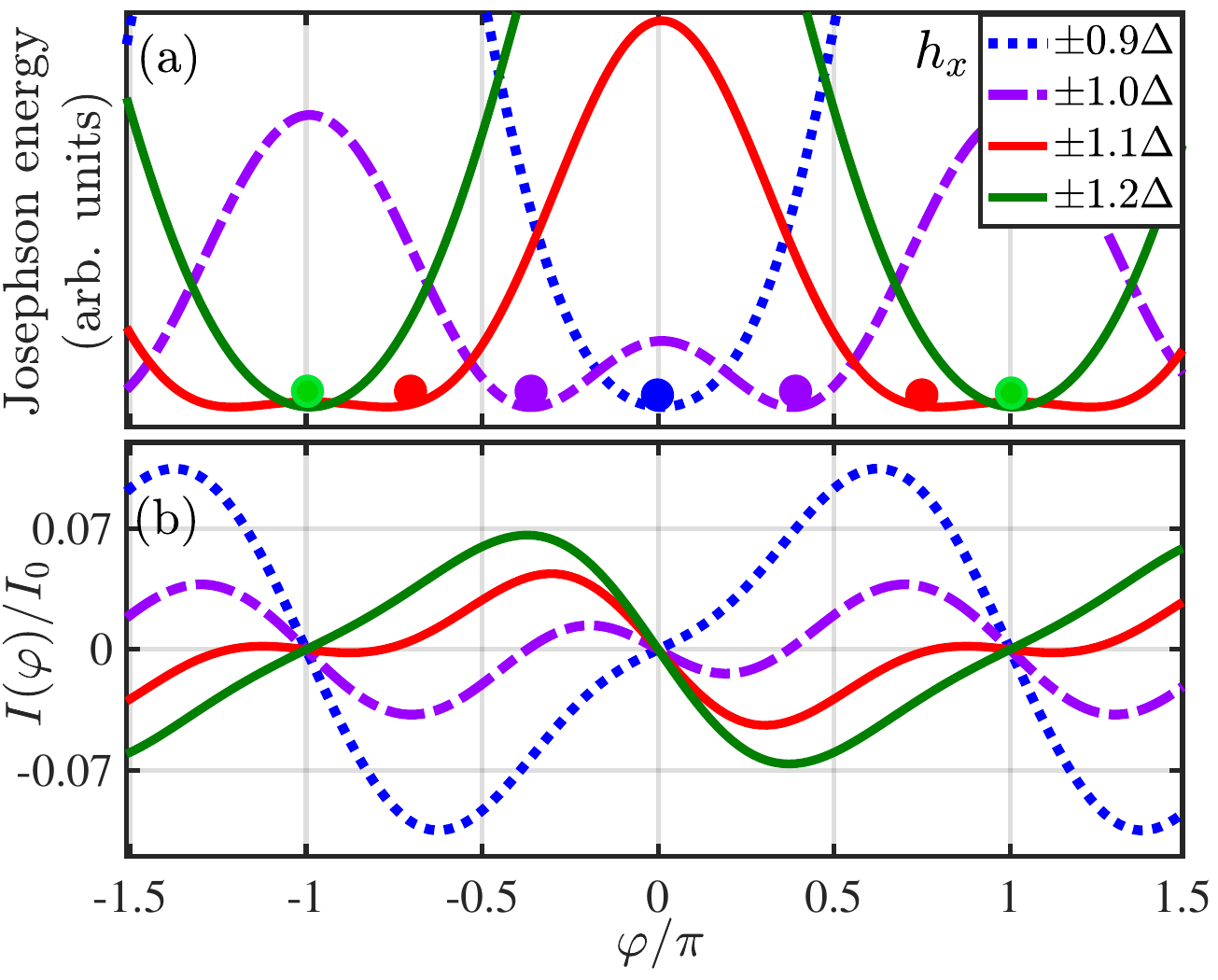}
\caption{(a) Josephson energy and (b) associated supercurrent evolution with the superconducting phase difference $\varphi$. 
Zeeman field values, $h_x$, are chosen near a $0$-$\pi$ transition.  
The other parameters are  $\ac=\pm0.1$ and $\bc=0$, $\mu=\Delta$, $h_y=0$.}
\label{fig2}
\end{figure}

The complexity of ${\cal H}$ precludes simple solutions and we evaluate the wavefunctions and supercurrent numerically.
To reduce the edge effects, we consider Fig.~\ref{fig1}(b) geometry with $W/d\gg 1$~\cite{Alidoust2015:JAP}.
This approach has been successfully used to study supercurrent in junctions with PSH, Weyl semimetals, and 
 phosphorene~\cite{Alidoust2020:PRB1,Alidoust2018:PRB1,Alidoust2020:PRB2,Alidoust2018:PRB2,Alidoust2018:PRB3}. The calculated 
 supercurrent 
 is normalized by $I_0=2|e\Delta|/\hbar$, where $e$ is the electron charge,  and $\Delta$ the energy gap in S. The energies are normalized by $\Delta$, lengths by $\xi_\text{S}=\hbar/\sqrt{2m^*\Delta}$, and cSOC strengths by $\Delta \xi_\text{S}^3$. The junction length is set at $d=0.3\xi_\text{S}$. 

To investigate the role of cSOC on the ground-state Josephson energy, $E_\text{GS}$, and the CPR obtained from the supercurrent $I(\varphi)\propto \partial E_\text{GS}/\partial \varphi$, we first
consider a simple situation with only Rashba cSOC ($\alpha_c\neq0$, $\beta_c=0$) and effective Zeeman field $h_x$ ($h_y=h_z=0$). 
The evolution of $E_\text{GS}$ with $|h_x|$, where its minima are denoted by dots in Fig.~\ref{fig2}(a), shows a continuous transition from $\varphi=0$ to $\pi$ state (blue to green dot). 
For $\varphi_0\neq0$,  $E_\text{GS}$ minima come in pairs at $\pm \varphi_0$~\cite{Sickinger2012:PRL}. 
The corresponding CPR reveals in Fig.~\ref{fig2}(b) a competition between the standard, $\sin \varphi$, and the next harmonic, 
$\sin 2\varphi$, resulting in $I(-\varphi)=-I(\varphi)$. There is no spontaneous current 
expected in a Josephson junction with SOC, $I(\varphi=0)=0$, but only $I_c$ reversal with $h_x$. 
Such a scenario of a continuous and symmetric 0-$\pi$ transition is well studied without SOC in S/ferromagnet/S JJs due to the changes in the effective magnetization or a thickness 
of the magnetic region~\cite{Kontos2002:PRL,Ryazanov2001:PRL,Bergeret2005:RMP,Eschrig2003:PRL,%
Halterman2015:PRB,Wu2018:PRB,Moen2020:PRB,Yokoyama2014:PRB}. 
\begin{figure}[t]
\centering
\includegraphics[width=0.465\textwidth]{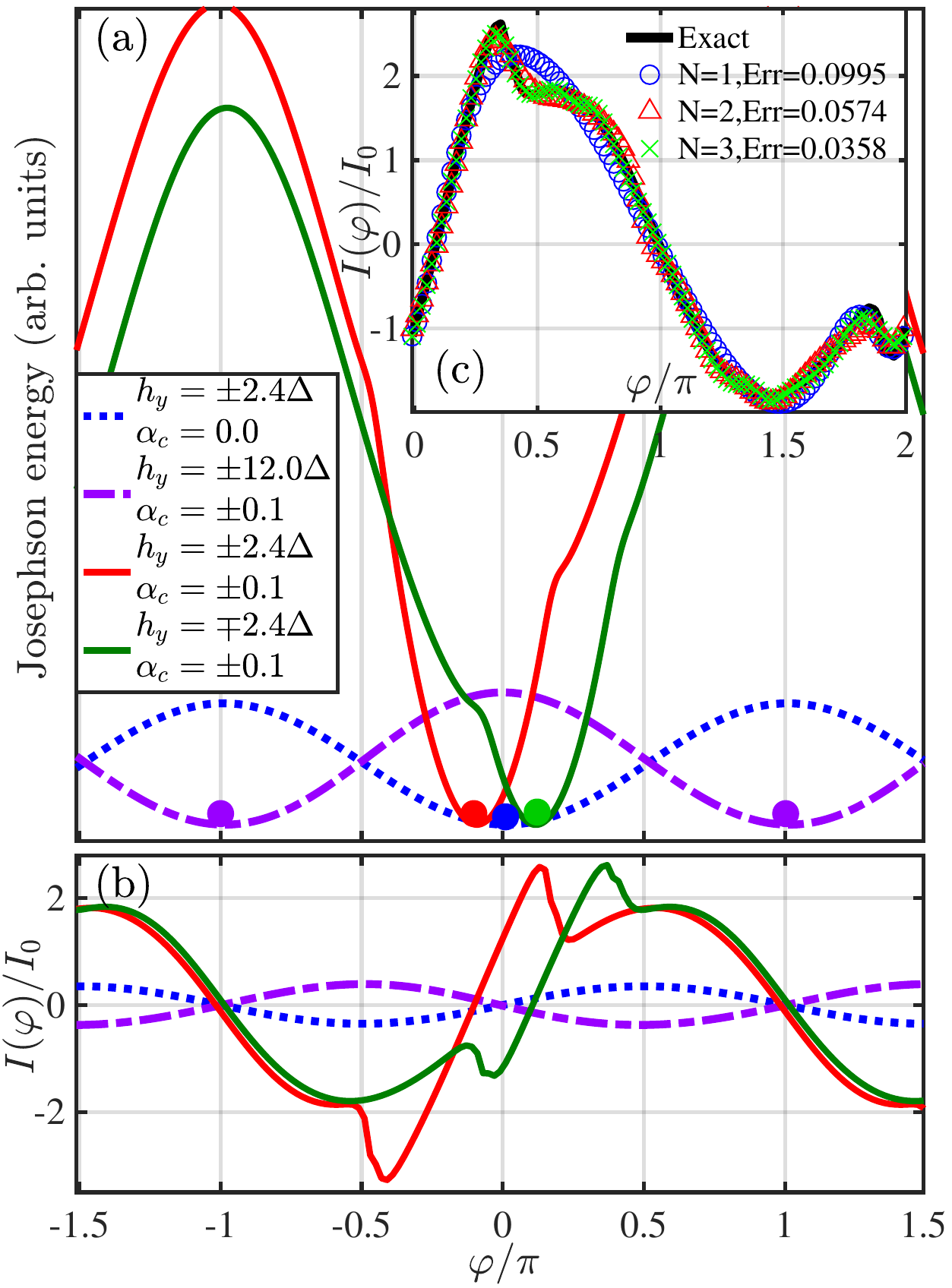}
\caption{(a) Josephson energy and (b) related supercurrent evolution with the superconducting phase difference $\varphi$
Zeeman field, $h_y$, at a fixed magnitude and varying Rashba cSOC strength $\ac$ are considered. 
The other parameters are $\bc=0$, $\mu=\Delta$, $h_x=0$. (c) Three fits to the green curve in (b) using the generalized CPR from Eq.~(\ref{cpr}) with
$N=1,2,3$ harmonics.}
\label{fig3}
\end{figure}

While our previous results suggest no direct cSOC influence on CPR, a simple in-plane rotation of {\bf h}, $h_x=0$, $h_y\neq0$, drastically changes this behavior.
This is shown in Figs.~\ref{fig3}(b) where, at fixed $|h_y|=2.4\Delta$, we see a peculiar influence of a finite Rashba cSOC which is responsible for the anomalous Josephson effect with spontaneous current, $I(\varphi=0)\neq0$,
and strong anharmonic CPR that cannot be described by $I(\varphi)=I_c\sin(\varphi+\varphi_0)$.
Unlike in Fig.~\ref{fig3}(a),  a relative sign between $\ac$ and $h$ alters the CPR and Josephson energy, where the ground states 
$\varphi_0$ appear at single points [green, red dots in Fig.~\ref{fig3}(a)], consistent with $\varphi_0\propto \alpha_c h_y$. 

If instead of $\mu=\Delta$, we consider a regime $\mu\gg \Delta$, the evolution of Josephson energy from Fig.~\ref{fig2}(a) changes.
While 0-$\pi$ transitions with $|h_x|$ remain, there are no longer global minima with $\varphi \neq 0,\pi$ and the CPR reveals a stronger anharmonicity. 
In contrast, for  $\mu\gg\Delta$, the anomalous Josephson effect from Fig.~\ref{fig3} remains robust and similar $\varphi_0$ states are accessible (see Ref.~\cite{sm}). 

Simple harmonics used to describe anharmonic CPR in high-temperature superconductors~\cite{Golubov2004:RMP,Kashiwaya2000:RPP}
here are not very suitable. 
By generalizing a short-junction limit for CPR~\cite{Yokoyama2014:PRB,Golubov2004:RMP,Hart2019:PRB}, we identify a much more compact form where only a small number of 
terms gives an accurate description. To recognize the importance of SOC and two nondegenerate spin channels, $\sigma$, we write
\begin{equation}\label{cpr}
I(\varphi) \approx \sum_{n=1}^N\sum_{\sigma=\pm}\frac{I_n^\sigma\sin(n\varphi+\varphi_{0n}^\sigma)}{\sqrt{1-\tau_n^\sigma\sin^2(n\varphi/2+\varphi_{0n}^\sigma/2)}},
\end{equation}
where  $\tau_n^\sigma$ is the normal region transparency for spin channel $\sigma$. With only few lowest terms in this expansion ($N=1,2,3$), 
shown in Fig.~\ref{fig3}(c) with the corresponding errors, it is possible to very accurately describe strong CPR anharmonicities for anomalous Josephson effect. 
To achieve the relative error from $N=3$ expansion in Eq.~(\ref{cpr}), in a standard $\{\sin,\cos\}$ expansion, with the corresponding 
phase shifts as extra fitting parameters, requires $N > 20$~\cite{sm}.  

Key insights into the CPR and an explicit functional dependence for the $\varphi_0$ state are obtained by a systematic $I(\varphi)$ symmetry analysis 
with respect to the cSOC ($\alpha_c$, $\beta_c$) and Zeeman field  or, equivalently, magnetization ($h_{x,y,z}$) parameters~\cite{sm}.
We find that $h_z$ plays no role in inducing the $\varphi_0$ state, it  
only produces $I(\varphi)$ reversals, 
explaining our focus on $h_z=0$ [Figs.~\ref{fig2} and \ref{fig3}]. 

These properties are expressed as an effective phase shift to the a sinusoidal CPR, $\sin(\varphi + \varphi_0)$, extracted from Eq.~(\ref{cpr}). 
We again distinguish small- and large-$\mu$ regime ($\mu=\Delta$ v.s. $\mu=10\Delta$). In the first case, for the JJ geometry 
from Fig.~\ref{fig1}, we obtain
\begin{equation}\label{phi_mu1}
\varphi_0 \propto \Gamma_y\Big( \alpha_\text{c}^2+\Gamma_1\beta_\text{c}^2\Big)h_x\bc+\Gamma_x\Big( \alpha_\text{c}^2-\Gamma_2\beta_\text{c}^2\Big)h_y\ac,
\end{equation}
where the 
parameters $\Gamma_{1,2,x,y}$ are introduced through their relations, $\Gamma_2>\Gamma_1$, $\Gamma_1<1$, $\Gamma_2>1$, $\Gamma_y(h_y=0)=\Gamma_x(h_x=0)=1$,  
$\Gamma_y(h_y\neq 0)<1$, $\Gamma_x(h_x\neq 0)< 1$. 
These relations are modified as $\mu$ and $\mathbf{h}$ change.
For $\mu \gg \Delta$, the functional dependence for the $\varphi_0$ state is simplified  
\begin{equation}
\varphi_0\propto \Big( \alpha_\text{c}^2-\Gamma_1\beta_\text{c}^2\Big)h_x\bc+\Big( \alpha_\text{c}^2-\Gamma_2\beta_\text{c}^2\Big)h_y\ac,
\label{phi_mu10}
\end{equation}
where  $\Gamma_2>\Gamma_1$ and $\Gamma_{1,2}>1$. 
Therefore, $\varphi_0$ state occurs when {\bf h} shifts {\bf p} $\bot$ to ${\bm I}(\varphi)$ and thus alters the SOC~\cite{sm}.

Taken together, these results reveal that cSOC in JJ supports a  large tunability of the Josephson energy, anharmonic CPR, and the anomalous phase, key
to many applications, from post-CMOS logic, superconducting spintronics, quiet qubits, and topological quantum computing. 
Realizing $\pi$ states in JJs is desirable for improving rapid single flux quantum (RSFQ) logic, with operation  
$>100\,$GHz~\cite{Likharev1991:IEEETAS,Terzioglu1998:IEEETAS}
and enhancing coherence by decoupling superconducting qubits from the environment~\cite{Yamashita2005:PRL}. 
However, common approaches for $\pi$ states using JJs combining $s$- and $d$-wave superconductors or  
JJs with ferromagnetic regions~\cite{Golubov2004:RMP,Kashiwaya2000:RPP} pose various limitations.   
Instead, extensively studied gate-tunable 
SOC~\cite{Zutic2004:RMP,Dettwiler2017:PRX,Nakamura2012:PRL,Moriya2014:PRL,Mayer2019:P,Nitta1997:PRL}, 
could allow not only a fast transformation
between $0$ and $\pi$ states in JJs with cSOC, but also an arbitrary $\varphi_0$ state to tailor desirable CPR.

An insight to the phase evolution and circuit operation of JJs with cSOC is provided by generalizing the classical model of
resistively and capacitively shunted junction (RSCJ)~\cite{Stewart1968:APL}. The total  current, $i$, is the sum of the displacement
current across the capacitance, $C$, normal current characterized by the resistance, $R$, and 
$I(\varphi)$,
\begin{equation}\label{pendulum}
\frac{\phi_0}{2\pi}C\frac{d^2\varphi}{dt^2} + \frac{\phi_0}{2\pi R}\frac{d\varphi}{dt} + I(\varphi)=i,
\end{equation}
where $\phi_0$ is the magnetic flux quantum and $I(\varphi)$ yields a generally anharmonic CPR, as shown from Eq.~(\ref{cpr}),
which can support $0$, $\pi$, and turnable $\varphi_0$ states.  As we have seen from Figs.~\ref{fig2} and \ref{fig3}, this CPR tunability is accompanied 
by the changes in Josephson energy, which in turn is responsible for the changes in effective values of $C$, $R$, and
the nonlinear Josephson inductance. This JJ tunability complements using voltage or flux control~\cite{Casparis2018:NN,Krantz2019:APR}. 
\begin{figure}[b]
\centering
\includegraphics[width=0.485\textwidth]{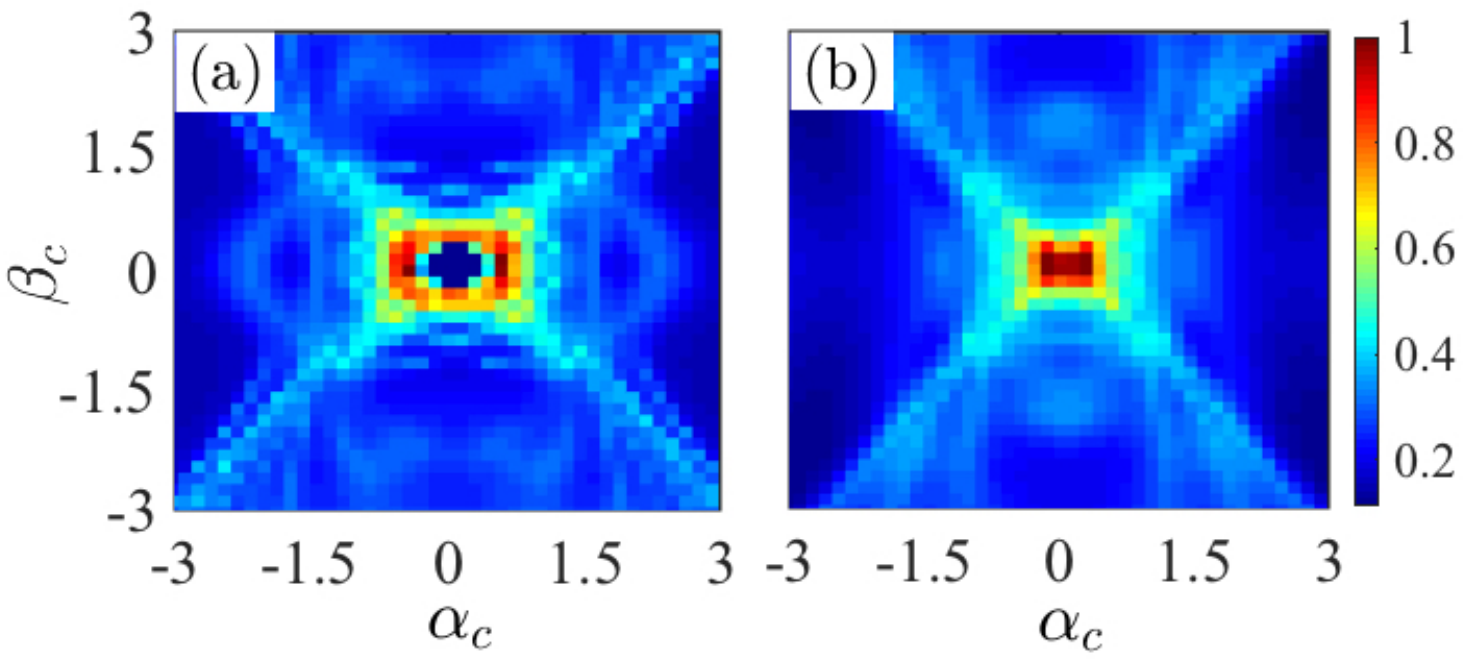}
\caption{\label{bs1} Normalized critical supercurrent as a function of cSOC strength $\ac$ and $\bc$ for (a) $\mu=\Delta$ and (b) $\mu=10\Delta$. The Zeeman field is set to zero.}
\label{fig4}
\end{figure}

In JJs with ferromagnetic regions,  $I_c$  is the tunable $I_c$ by changing the underlying magnetic state~\cite{Gingrich2016:NP,Baek2014:NC,Costa2017:PRB}.
In JJs with cSOC, tuning $I_c$ could be realized through gate control by changing the relative strengths of $\alpha_c$ and $\beta_c$,
even at zero Zeeman field. This is shown in Fig.~\ref{fig4} by calculating $\text{Max}[I(\varphi)]$ with $\varphi\in [0,2\pi]$.  
In the low-$\mu$ regime, the maximum $I_c$ occurs at the slightly curved region near the symmetry lines $|\alpha_c|=|\beta_c|$. 
For the high-$\mu$ regime, the region of maximum $I_c$ evolves into inclined symmetry lines,  $|\alpha_c|={\cal A}|\beta_c|$, ${\cal A} <1$. 
Similar to linear SOC, in the diffusive regime for cSOC, one expects that the minimum in $I_c$ occurs near these symmetry lines because of the presence of long-range spin-triplet supercurrent~\cite{Alidoust2015:NJP,Alidoust2020:PRB1}.  
 
\begin{figure}[t]
\centering
\includegraphics[width=0.47\textwidth]{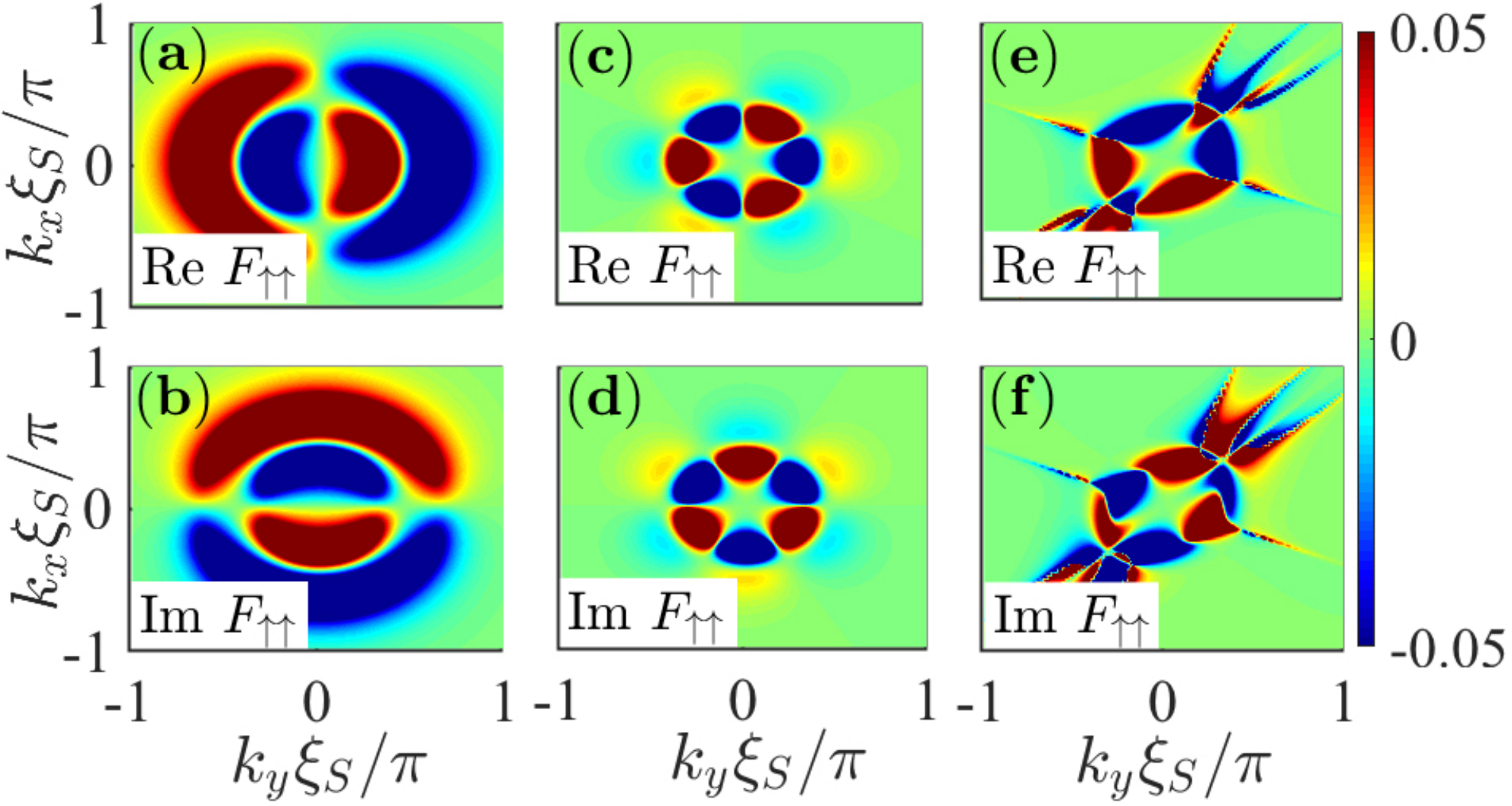}
\caption{Real and imaginary parts of equal-spin superconducting correlations in the k-space, $\xi_\text{S}=\hbar/\sqrt{2m^*\Delta}$ is the characteristic length.  
(a), (b) Linear Rashba, $\alpha=1$. (c), (d) cSOC, $\alpha_c=1$, $\beta_c=0$. (e), (f) cSOC, $\alpha_c=\beta_c=1$. The other parameters are the same for all panels.}
\label{fig5}
\end{figure}

We expect that a hallmark of JJs with cSOC goes beyond CPR and will also influence the spin structure and symmetry properties of superconducting proximity effects.
Linear SOC is responsible for mixed singlet-triplet superconducting pairing~\cite{Gorkov2001:PRL}, while  with Zeeman or exchange field it is possible to favor spin-triplet proximity effects 
which can become long-range~\cite{Eschrig2015:RPP,Linder2015:NP} or host Majorna bound states~\cite{Lutchyn2010:PRL,Oreg2010:PRL}. 
To explore the proximity effects in the cSOC region, we  calculate superconducting pair correlations using the Matsubara representation 
for the anomalous Green function, $F(\tau; \mathbf{r},\mathbf{r}')$~\cite{Zagoskin:2014}, 
\begin{equation}
F_{ss'}(\tau; \mathbf{r},\mathbf{r}') = +\langle T_{\tau} \psi_s(\tau,\mathbf{r}) \psi_{s_1}(0,\mathbf{r}') \rangle (-i\sigma^y_{s_1 s'} ), 
\label{GF_F1}
\end{equation}
where $s, s', s_1$ are spin indices, the summation is implied over $s_1$, $\tau$ is the imaginary time, $\psi_s$ is the field operator, and $T_{\tau}$ denotes time ordering of operators~\cite{sm}.

For a translationally invariant SOC region, spin-triplet correlations in Fig.~\ref{fig5}, obtained from Eq.~(\ref{GF_F1}), provide a striking difference between linear and cubic SOC. Unlike  
the $p$-wave symmetry for linear Rashba SOC [Figs.~\ref{fig5}(a), \ref{fig5}(b)], we  see that the $f$-wave symmetry is the fingerprint for cSOC, retained with only 
$\alpha_c\neq0$ [Figs.~\ref{fig5}(c), \ref{fig5}(d)] or both $\alpha_c, \beta_c \neq 0$ [Figs.~\ref{fig5}(e), \ref{fig5}(f)]. Remarkably, unlike the commonly-sought $p$-wave symmetry, we confirm that with a suitable 
orientation of the Zeeman field cSOC also supports Majorana flat bands~\cite{sm}. 

While we are not aware of any Josephson effect experiments in 2D systems dominated by cSOC, our studied parameters are within the range of already reported measurements.
Choosing $m^*$ of an electron mass, and $\Delta=0.2\,$meV, which is similar for both Al and proximity-induced superconductivity~\cite{Mayer2019:P,Mayer2020:NC}, 
the characteristic length becomes $\xi_\text{S} \approx 14\,$nm. The resulting cSOC strength from Fig.~\ref{fig3}(b) with $\alpha_c \Delta \xi_\text{S}^3 \approx 50\,$eV\AA$^3$ is compatible with the values in 
2D electron and hole gases~\cite{Cottier2020:PRB, Liu2018:PRL}.  The Zeeman splitting $2.4 \times 0.2\,$meV is available by applying magnetic field in large $g$-factor materials~\cite{Zutic2004:RMP}, or from magnetic proximity effects, measured in 2D systems to reach up to $\sim 20\,$meV~\cite{Zutic2019:MT}. Even though we have mostly
focussed on the tunable Rashba SOC, the Dresselhaus SOC can also be gate tunable~\cite{Dettwiler2017:PRX,Iordanskii1994:JETPL}, offering a further control of 
the anomalous Josephson effect.

Our results reveal that the cSOC in JJs provides versatile opportunities to design a superconducting response
and test its unexplored manifestations. The anomalous Josephson effect could serve as a sensitive probe to quantify cSOC. 
While identifying the relevant form of SOC is a challenge even in the normal state~\cite{Zutic2004:RMP,Fabian2007:APS}, 
in the superconducting state already a modest SOC can give a strong anisotropy in the transport 
properties~\cite{Hogl2015:PRL,Martinez2020:PRA,Gonzalez-Ruano2020:PRB,Vezin2020:PRB}
and enable extracting the resulting SOC. Identifying SOC, either intrinsic, or generated through magnetic textures, remains important for understanding which systems 
could host Majorana bound 
states~\cite{Desjardins2019:NM,Scharf2019:PRB,Pakizer2020:P,Fatin2016:PRL,Matos-Abiague2017:SSC,Ronetti2020:PRR,Klinovaja2012:PRL,
Zhou2019:PRB,Mohanta2019:PRA,Turcotte2020:PRB,Jiang2021:N,Rex2020:PRB,Kornich2020:PRB,Mohanta2020:P,Mohanta2018:PRB}.

With the advances in gate-tunable structures 
and novel materials systems~\cite{Mayer2019:P,Mayer2020:NC,Nakamura2012:PRL,Moriya2014:PRL,Cottier2020:PRB,Liu2018:PRL,Assouline2019:NC}, 
the functional dependence of the anomalous phase $\varphi_0$ and the $f$-wave superconducting correlations 
could also enable decoupling of the linear and cubic SOC contributions~\cite{sm}. 
For the feasibility of such decoupling, it would be useful to consider methods employed in the studies of the nonlinear Meissner 
effect~\cite{Xu2015:PRB,Bae2019:RSI,Zhuravel2013:PRL,Prozorov2006:SST,Halterman2000:PRB,Bhattacharya1999:PRL,Zutic1997:PRB,Zutic1998:PRB}. 
Even small corrections to the supercurrent from the magnetic anisotropy of the nonlinear Meissner response offer a sensitive probe to distinguish different 
pairing-state symmetries.

\acknowledgments
M.A. was supported by Iran's National Elites Foundation (INEF). C.S. and I.\v{Z}. were supported by NSF ECCS-1810266, and I.\v{Z} by DARPA DP18AP900007, and the UB Center for Computational Research.

\appendix

\onecolumngrid

\section{Supplemental Material for: \\ Cubic spin-orbit coupling and anomalous Josephson effect in planar junctions} 

\section{I. Wave functions and eigenvalues} 

Here we present the four-component spinor wave functions for spin-orbit coupling cubic in the wave vector (cSOC) considered for the geometry of a Josephson junction discussed in the main text. 
To simplify the notation, we introduce the parameterization $k_x=k\cos\theta$, $k_y=k\sin\theta$ for the in-plane components of the wave vector, where $\theta$ is the incident angle of moving particles with respect to the interface.
 The out-of-plane component of Zeeman field is set to zero $h_z=0$ and the in-plane components are given by $ h_x=h_0\cos\phi, h_y=h_0\sin\phi$. 
 By diagonalizing the Hamiltonian in the main text, we find the following spinor 
 wave functions for the electron ($\psi_e$) and hole ($\psi_h$) parts,
\begin{subequations}
  \begin{equation}
  \psi_e^\pm = (\frac{\varepsilon_e^\pm-\hbar^2k^2/(2m^*)+\mu}{h_0 e^{+i\phi}-\beta_\text{c} k^3 e^{+i\theta}-i\alpha_\text{c} k^3 e^{+3i\theta}},1,0,0)^\text{T} e^{\pm i \mathbf{k}\cdot\mathbf{r}},
  \end{equation}
  
\begin{equation}
  \psi_h^\pm = (0,0,\frac{-\varepsilon_h^\pm -\hbar^2k^2/(2m^*)+\mu}{h_0e^{-i\phi}+\beta_\text{c} k^3 e^{-i\theta}-i\alpha_\text{c} k^3 e^{-3i\theta}},1)^\text{T} e^{\pm i \mathbf{k}\cdot\mathbf{r}},
  \end{equation}  
 \end{subequations} 
  
\begin{subequations}  
  \begin{equation}
 \varepsilon_e^\pm=+\hbar^2k^2/(2m^*)-\mu \pm \sqrt{h_0^2+(\alpha_\text{c}^2+\beta_\text{c}^2)k^6-2\big[+h_0\beta_\text{c}\cos(\theta-\phi)+\alpha_\text{c}\beta_\text{c} k^3\sin(2\theta)- h_0\alpha_\text{c}\sin(3\theta-\phi)\big ] k^3},
 \end{equation} 
 \begin{equation}
 \varepsilon_h^\pm=-\hbar^2k^2/(2m^*)+\mu \pm \sqrt{h_0^2+(\alpha_\text{c}^2+\beta_\text{c}^2)k^6-2\big[-h_0\beta_\text{c}\cos(\theta-\phi)+\alpha_\text{c}\beta_\text{c} k^3\sin(2\theta)+ h_0\alpha_\text{c}\sin(3\theta-\phi)\big] k^3},
 \end{equation} 
 \end{subequations} 
 where $\varepsilon_e$  ($\varepsilon_h$) are the electron (hole) quasiparticle energies, $m^*$ is the effective mass, 
 $\mu$ is the chemical potential, while $\alpha_c$ and $\beta_c$ are the Rashba and Dresselhaus cSOC strengths. The total wave function inside the middle region of the Josephson junction is a superposition of right and left moving electron and hole wave functions. The wave functions within the superconducting regions are the usual $s$-wave superconducting wave functions ($\pm$). In the calculations that follows, the energies are normalized by the superconducting gap $\Delta$, 
lengths by $\xi_\text{S}=\hbar/\sqrt{2m^*\Delta}$, and the cSOC strengths by $\Delta \xi_\text{S}^3$. 
  
\begin{figure*}
\centering
\includegraphics[width=\textwidth]{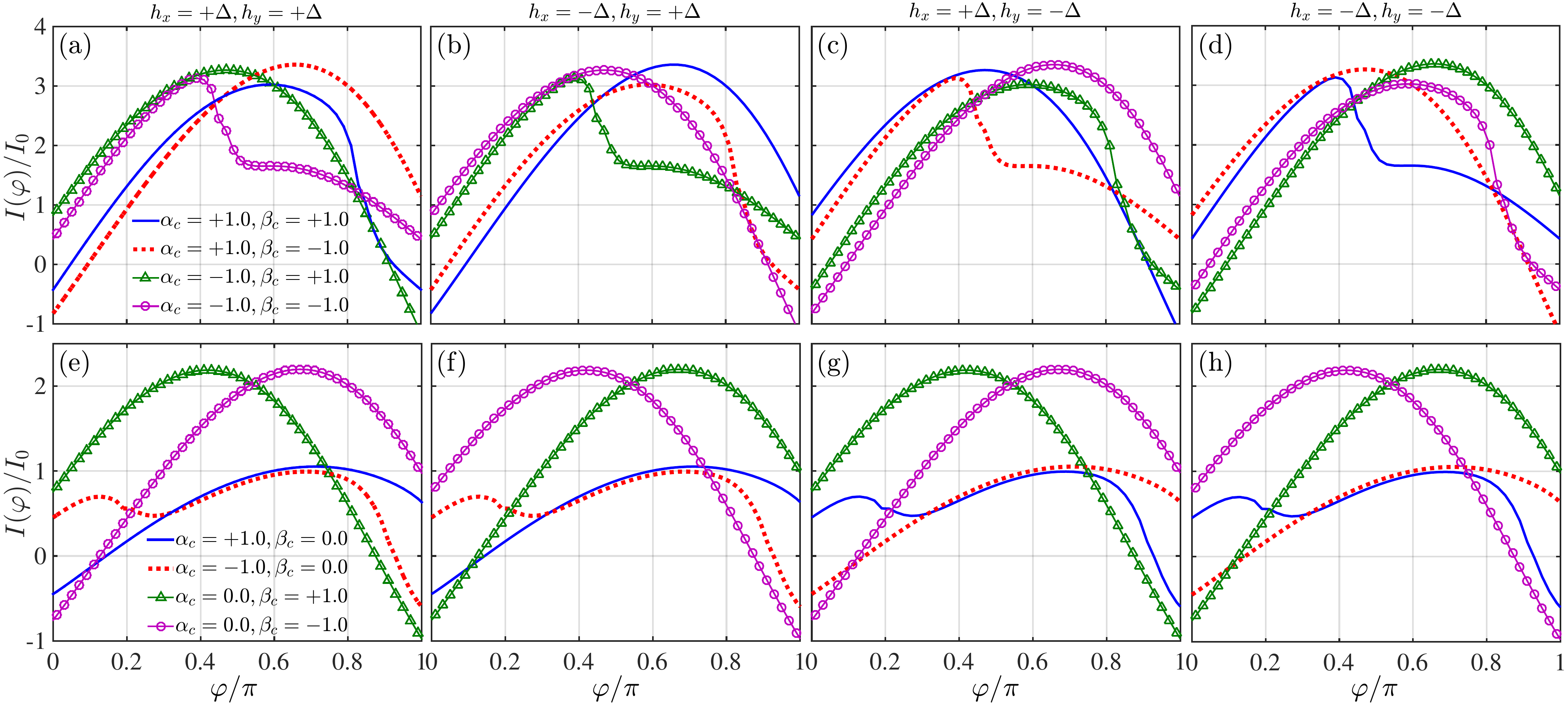}
\caption{\label{bs1} The current-phase relations where the chemical potential is fixed at $\mu=\Delta$. The components of Zeeman field change column-wise from the left column to the right column; (a), (e) $h_x=+h_y=+\Delta$, (b), (f) $h_x=-h_y=-\Delta$, (c), (g)  $-h_x=h_y=-\Delta$, and (d), (h) $h_x=h_y=-\Delta$. In the top and bottom row panels, the cSOC coefficients change; (a)-(d) $\ac=+1.0, \bc=\pm 1.0$, $\ac=-1.0, \bc=\pm 1.0$, (e)-(h) $\ac=0.0, \bc=\pm 1.0$, $\ac=\pm 1.0, \bc=0$.}\label{I_phi_smpl}
\end{figure*}

\section{II. Analysis of current-phase relations and obtaining the $\varphi_0$ state}

To obtain explicit functional form for the state of an anomalous superconducting phase difference in the Josephson junction,  $\varphi_0$, in both low- and high-chemical potential regimes, we have computed current-phase relations (CPRs) at $\mu=\Delta$, and $\mu=10\Delta$. A sample plot, of numerous plots produced is shown in Fig.~\ref{I_phi_smpl}, where the supercurrent, $I$, normalized by $I_0=2|e\Delta|/\hbar$, is plotted against the superconducting phase difference $\varphi$ for four different sets of the Zeeman field components $h_x=\pm\Delta, h_y=+\Delta$ and $h_x=\pm\Delta, h_y=-\Delta$, corresponding to a column each. In Figs.~\ref{I_phi_smpl}(a)-\ref{I_phi_smpl}(d), we consider $\ac=+1,\bc=\pm 1$ and $\ac=-1,\bc=\pm 1$ whereas in Figs.~\ref{I_phi_smpl}(e)-\ref{I_phi_smpl}(h), we set $\ac=\pm 1,\bc=0$ and $\ac=0,\bc=\pm 1$. As seen, the supercurrent at $\varphi=0$ is nonzero for all parameter value sets considered. To obtain the functional form for the $\varphi_0$ state, we have conducted a systematic study of CPRs. The results at $\mu=\Delta$ are summarized in Eqs.~(\ref{I1_set1})-(\ref{I1_set5}) and the status of the $\varphi_0$ state is given in front of CPRs;
\begin{subequations}\label{I1_set1}
\begin{eqnarray}
&& I(+\alpha_\text{c},0,\pm h_x,0)=I(-\alpha_\text{c},0,\pm h_x,0);\; \varphi_0=0,\\&& 
I(0,+\beta_\text{c},0,\pm h_y)=I(0,- \beta_\text{c},0,\pm h_y);\; \varphi_0=0,
\end{eqnarray}
\end{subequations}
\begin{subequations}\label{I1_set2}
\begin{eqnarray}
&& I(+\alpha_\text{c},0,0,+ h_y)=I(-\alpha_\text{c},0,0,- h_y);\; \varphi_0>0,\label{s2_1}\\
&& I(+\alpha_\text{c},0,0,- h_y)=I(-\alpha_\text{c},0,0,+ h_y);\; \varphi_0<0,\label{s2_2}\\
&& I(0,+\beta_\text{c}, + h_x, 0)=I(0,-\beta_\text{c}, - h_x, 0);\; \varphi_0>0,\label{s2_3}\\
&& I(0,+\beta_\text{c}, - h_x, 0)=I(0,-\beta_\text{c}, + h_x, 0);\; \varphi_0<0.\label{s2_4}
\end{eqnarray}
\end{subequations}
The relations in Eqs.~(\ref{I1_set1}) and (\ref{I1_set2}) offer a function of type $\varphi_0\propto {\cal O}(h_x){\cal O}(\beta_\text{c}) + {\cal O}(h_y){\cal O}(\alpha_\text{c})$ for the $\varphi_0$ state. The $\cal O$ functions are odd and unequal unless otherwise stated. We have introduced them to simplify our analysis. The visualized CPRs illustrate that the modulus of the $\varphi_0$ state is the same in Eqs.~(\ref{s2_1}) and (\ref{s2_2}) as well as in Eqs.~(\ref{s2_3}) and (\ref{s2_4}). In addition, the modulus of the $\varphi_0$ state in Eqs. (\ref{s2_1}) and (\ref{s2_2}) is larger than that of Eqs. (\ref{s2_3}) and (\ref{s2_4}).  
\begin{subequations}\label{I1_set3}
\begin{align}
& I(+\alpha_\text{c},+\beta_\text{c}, 0, +h_y)=I(+\alpha_\text{c},-\beta_\text{c}, 0, +h_y)= I(-\alpha_\text{c},+\beta_\text{c}, 0, -h_y)=I(-\alpha_\text{c},-\beta_\text{c}, 0, -h_y);\; \varphi_0<0,\label{s3_3}\\
& I(-\alpha_\text{c},+\beta_\text{c}, 0, +h_y)=I(-\alpha_\text{c},-\beta_\text{c}, 0, +h_y)= I(+\alpha_\text{c},+\beta_\text{c}, 0, -h_y)=I(+\alpha_\text{c},-\beta_\text{c}, 0, -h_y);\; \varphi_0>0,\label{s3_4}\\
& I(+\alpha_\text{c},+\beta_\text{c}, + h_x, 0)=I(-\alpha_\text{c},+\beta_\text{c}, + h_x, 0)=
I(+\alpha_\text{c},-\beta_\text{c}, - h_x, 0)=I(-\alpha_\text{c},-\beta_\text{c}, - h_x, 0);\; \varphi_0>0,\label{s3_1}\\
& I(+\alpha_\text{c},-\beta_\text{c}, + h_x, 0)=I(-\alpha_\text{c},-\beta_\text{c}, + h_x, 0)= I(+\alpha_\text{c},+\beta_\text{c}, - h_x, 0)=I(-\alpha_\text{c},+\beta_\text{c}, - h_x, 0);\; \varphi_0<0,\label{s3_2}
\end{align}
\end{subequations}
The relations in Eqs.~(\ref{s3_3}) and (\ref{s3_4}) show that the nonzero $\beta_\text{c}$ reverses the sign of the $\varphi_0$ state compared to Eqs.~(\ref{s2_1}) and (\ref{s2_2}) whereas the inclusion of $\alpha_\text{c}$ in 
Eqs.~(\ref{s3_1}) and (\ref{s3_2}) leaves the sign of the $\varphi_0$ state intact compared to Eqs.~(\ref{s2_3}) and (\ref{s2_4}). Furthermore, the modulus of the $\varphi_0$ state in Eqs.~(\ref{s3_1}) and (\ref{s3_2}) is the same. The same statement is true for Eqs.~(\ref{s3_3}) and (\ref{s3_4}) although the modulus of $\varphi_0$ in Eqs.~(\ref{s3_1}) and (\ref{s3_2}) is larger than 
Eqs.~(\ref{s3_3}) and (\ref{s3_4}). The analysis so far offers $\varphi_0\propto (\alpha_\text{c}^2+\Gamma_1\beta_\text{c}^2)h_x\beta + (\alpha_\text{c}^2-\Gamma_2\beta_\text{c}^2)h_y\alpha$, in which we have defined the constants $\Gamma_2>\Gamma_1$, $\Gamma_1<1$, and $\Gamma_2>1$. 
\begin{subequations}\label{I1_set4}
\begin{eqnarray}
&& I(+\alpha_\text{c},0, + h_x, +h_y)=I(-\alpha_\text{c},0, + h_x, -h_y)=
I(+\alpha_\text{c},0, - h_x, +h_y)=I(-\alpha_\text{c},0, - h_x, -h_y); \; \varphi_0<0,\label{s4_1}\\
&& I(-\alpha_\text{c},0, + h_x, +h_y)=I(+\alpha_\text{c},0, + h_x, -h_y)=
I(-\alpha_\text{c},0, - h_x, +h_y)=I(+\alpha_\text{c},0, - h_x, -h_y); \; \varphi_0>0,\label{s4_2}\\
&& I(0,-\beta_\text{c}, + h_x, +h_y)=I(0,-\beta_\text{c}, + h_x, -h_y)=
I(0,+\beta_\text{c}, - h_x, +h_y)=I(0,+\beta_\text{c}, - h_x, -h_y);\; \varphi_0<0,\label{s4_3}\\
&& I(0,+\beta_\text{c}, + h_x, +h_y)=I(0,+\beta_\text{c}, + h_x, -h_y)=
I(0,-\beta_\text{c}, - h_x, +h_y)=I(0,-\beta_\text{c}, - h_x, -h_y);\; \varphi_0>0. \label{s4_4} 
\end{eqnarray}
\end{subequations}
To understand the influence of Zeeman field on the inferred $\varphi_0$ so far, we have computed CPRs with $\alpha_\text{c}=0, \beta_\text{c}=\pm 1$ and $\alpha_\text{c}=\pm, \beta_\text{c}=0$ in the presence of $h_x\neq 0$ and $h_y\neq 0$. The results are given by Eqs.~(\ref{I1_set4}). As it can be seen, the inclusion of the second component of Zeeman field into Eqs.~(\ref{I1_set2}) is unable to change the sign of $\varphi_0$. Nevertheless, the modulus of the $\varphi_0$ states in the absence of the second component of Zeeman field is larger than the case when the second component is incorporated. Therefore, some coefficients dependent on the second component of Zeeman field are required. Considering the findings above, the $\varphi_0$ state takes the following form 
\begin{equation}\label{phi_mu1}
\varphi_0\propto \Gamma_y\Big( \alpha_\text{c}^2+\Gamma_1\beta_\text{c}^2\Big)h_x\beta_\text{c}+\Gamma_x\Big( \alpha_\text{c}^2-\Gamma_2\beta_\text{c}^2\Big)h_y \alpha_\text{c},
\end{equation}
in which the parameters $\Gamma_{1,2,x,y}$ are introduced through their relations: $\Gamma_2>\Gamma_1$, $\Gamma_1<1$, $\Gamma_2>1$, $\Gamma_y(h_y=0)=\Gamma_x(h_x=0)=1$ and $\Gamma_y(h_y\neq 0)<1, \Gamma_x(h_x\neq 0)< 1$. To complete the analysis, we have computed CPRs in the presence of both cSOC coefficients $ \alpha_\text{c}, \beta_\text{c}$ and Zeeman field or, equivalently, magnetization components $h_x,h_y$. The resulting CPRs are presented in Eqs.~(\ref{I1_set5}), 
\begin{subequations}\label{I1_set5}
\begin{eqnarray}
&& I(+\alpha_\text{c},-\beta_\text{c}, +h_x,+ h_y)=I(+\alpha_\text{c},+\beta_\text{c}, -h_x,+ h_y)=
I(-\alpha_\text{c},-\beta_\text{c}, +h_x,- h_y)=I(-\alpha_\text{c},+\beta_\text{c}, -h_x,- h_y);\; \varphi_0<0,\\
&& I(+\alpha_\text{c},+\beta_\text{c}, +h_x,+ h_y)=I(+\alpha_\text{c},-\beta_\text{c}, -h_x,+ h_y)=
I(-\alpha_\text{c},+\beta_\text{c}, +h_x,- h_y)=I(-\alpha_\text{c},-\beta_\text{c}, -h_x,- h_y);\; \varphi_0<0,\\
&& I(-\alpha_\text{c},-\beta_\text{c}, +h_x,+ h_y)=I(-\alpha_\text{c},+\beta_\text{c}, -h_x,+ h_y)=
I(+\alpha_\text{c},-\beta_\text{c}, +h_x,- h_y)=I(+\alpha_\text{c},+\beta_\text{c}, -h_x,- h_y);\; \varphi_0>0,\\
&& I(-\alpha_\text{c},+\beta_\text{c}, +h_x,+ h_y)=I(-\alpha_\text{c},-\beta_\text{c}, -h_x,+ h_y)=
I(+\alpha_\text{c},+\beta_\text{c}, +h_x,- h_y)=I(+\alpha_\text{c},-\beta_\text{c}, -h_x,- h_y);\; \varphi_0>0.
\end{eqnarray}
\end{subequations}
As can be seen, the relations above confirm the obtained spontaneous (at $I(\varphi=0)$) phase-shift given by Eq.~(\ref{phi_mu1}).

We have carried out the same calculations and analysis described so far for the case of a high-chemical potential regime, i.e., $\mu=10\Delta$. The CPRs obtained are summarized in Eqs.~(\ref{I10_set1})-(\ref{I10_set5}).
\begin{subequations}\label{I10_set1}
\begin{eqnarray}
&& I(+\alpha_\text{c},0,\pm h_x,0)=I(-\alpha_\text{c},0,\pm h_x,0);\; \varphi_0=0,\\&&
I(0,+\beta_\text{c},0,\pm h_y)=I(0,- \beta_\text{c},0,\pm h_y);\; \varphi_0=0,
\end{eqnarray}
\end{subequations}
\begin{subequations}\label{I10_set2}
\begin{eqnarray}
&& I(+\alpha_\text{c},0,0,+ h_y)=I(-\alpha_\text{c},0,0,- h_y);\; \varphi_0>0,\label{s10_21}\\
&& I(+\alpha_\text{c},0,0,- h_y)=I(-\alpha_\text{c},0,0,+ h_y);\; \varphi_0<0,\label{s10_22}\\
&& I(0,+\beta_\text{c}, + h_x, 0)=I(0,-\beta_\text{c}, - h_x, 0);\; \varphi_0<0,\label{s10_23}\\
&& I(0,+\beta_\text{c}, - h_x, 0)=I(0,-\beta_\text{c}, + h_x, 0);\; \varphi_0>0.\label{s10_24}
\end{eqnarray}
\end{subequations}
Compared with Eqs.~(\ref{I1_set2}), from the above results, the $\varphi_0$ state has changed sign and therefore, offers a relation of the type $\varphi_0\propto {\cal O}(h_x){\cal O}(\beta_\text{c}) - {\cal O}(h_y){\cal O}(\alpha_\text{c})$. We note that the modulus of the $\varphi_0$ phase shift remains the same in Eqs.~(\ref{s10_21}) and (\ref{s10_22}) and in Eqs.~(\ref{s10_23}) and (\ref{s10_24}). However, the modulus of the phase shift in Eqs.~(\ref{s10_21}), (\ref{s10_22}) is smaller than that of Eqs.~(\ref{s10_23}), (\ref{s10_24}).
\begin{subequations}\label{I10_set3}
\begin{align}
& I(+\alpha_\text{c},+\beta_\text{c}, 0, +h_y)=I(+\alpha_\text{c},-\beta_\text{c}, 0, +h_y)= I(-\alpha_\text{c},+\beta_\text{c}, 0, -h_y)=I(-\alpha_\text{c},-\beta_\text{c}, 0, -h_y);\; \varphi_0<0,\label{s10_31}\\
& I(-\alpha_\text{c},+\beta_\text{c}, 0, +h_y)=I(-\alpha_\text{c},-\beta_\text{c}, 0, +h_y)= I(+\alpha_\text{c},+\beta_\text{c}, 0, -h_y)=I(+\alpha_\text{c},-\beta_\text{c}, 0, -h_y);\; \varphi_0>0,\label{s10_32}\\
& I(+\alpha_\text{c},+\beta_\text{c}, + h_x, 0)=I(-\alpha_\text{c},+\beta_\text{c}, + h_x, 0)=
I(+\alpha_\text{c},-\beta_\text{c}, - h_x, 0)=I(-\alpha_\text{c},-\beta_\text{c}, - h_x, 0);\; \varphi_0<0,\label{s10_33}\\
& I(+\alpha_\text{c},-\beta_\text{c}, + h_x, 0)=I(-\alpha_\text{c},-\beta_\text{c}, + h_x, 0)= I(+\alpha_\text{c},+\beta_\text{c}, - h_x, 0)=I(-\alpha_\text{c},+\beta_\text{c}, - h_x, 0);\; \varphi_0>0.\label{s10_34}
\end{align}
\end{subequations}
By setting nonzero $\beta_\text{c}$ and $\alpha_\text{c}$ in Eqs.~(\ref{s10_21}), (\ref{s10_22}) and Eqs.~(\ref{s10_23}), (\ref{s10_24}), respectively, the main findings are summarized in 
Eqs.~(\ref{I10_set3}). The modulus of the phase shift is the same in Eqs.~(\ref{s10_31}) and (\ref{s10_32}). The same observation is true in Eqs.~(\ref{s10_33}) and (\ref{s10_34}). However, the modulus of the $\varphi_0$ state in Eqs.~(\ref{s10_31}), (\ref{s10_32}) is smaller than that of Eqs.~(\ref{s10_33}), (\ref{s10_34}). Therefore, one can conclude a relation for the phase shift as $\varphi_0\propto (\alpha_\text{c}^2-\Gamma_1\beta_\text{c}^2)h_x\beta + (\alpha_\text{c}^2-\Gamma_2\beta_\text{c}^2)h_y\alpha$ in which the constants follow $\Gamma_2>\Gamma_1$ and $\Gamma_{1,2}>1$. 
\begin{subequations}\label{I10_set4}
\begin{eqnarray}
&& I(-\alpha_\text{c},0, + h_x, +h_y)=I(+\alpha_\text{c},0, + h_x, -h_y)=
I(-\alpha_\text{c},0, - h_x, +h_y)=I(+\alpha_\text{c},0, - h_x, -h_y); \; \varphi_0<0,\\
&& I(+\alpha_\text{c},0, + h_x, +h_y)=I(-\alpha_\text{c},0, + h_x, -h_y)=
I(+\alpha_\text{c},0, - h_x, +h_y)=I(-\alpha_\text{c},0, - h_x, -h_y); \; \varphi_0>0,\\
&& I(0,+\beta_\text{c}, + h_x, +h_y)=I(0,+\beta_\text{c}, + h_x, -h_y)=
I(0,-\beta_\text{c}, - h_x, +h_y)=I(0,-\beta_\text{c}, - h_x, -h_y);\; \varphi_0<0,\\
&& I(0,-\beta_\text{c}, + h_x, +h_y)=I(0,-\beta_\text{c}, + h_x, -h_y)=
I(0,+\beta_\text{c}, - h_x, +h_y)=I(0,+\beta_\text{c}, - h_x, -h_y);\; \varphi_0>0.
\end{eqnarray}
\end{subequations}
To investigate the influence of the components of Zeeman field on the $\varphi_0$ state, we have added the second component of Zeeman field in Eqs.~(\ref{I10_set2}). In Eqs.~(\ref{I10_set4}), we have found that the inclusion of the second component of Zeeman field induces no sign and modulus changes to the $\varphi_0$ state compared with Eqs.~(\ref{I10_set2}). Therefore, the latest functional form of the $\varphi_0$ state above is valid without further modifications.  
\begin{subequations}\label{I10_set5}
\begin{eqnarray}
&& I(+\alpha_\text{c},+\beta_\text{c}, +h_x,+ h_y)=I(-\alpha_\text{c},+\beta_\text{c}, +h_x,- h_y)=
I(+\alpha_\text{c},-\beta_\text{c}, -h_x,+ h_y)=I(-\alpha_\text{c},-\beta_\text{c}, -h_x,- h_y);\; \varphi_0<0,\label{s10_51}\\
&& I(+\alpha_\text{c},-\beta_\text{c}, +h_x,+ h_y)=I(-\alpha_\text{c},-\beta_\text{c}, +h_x,- h_y)=
I(+\alpha_\text{c},+\beta_\text{c}, -h_x,+ h_y)=I(-\alpha_\text{c},+\beta_\text{c}, -h_x,- h_y);\; \varphi_0<0,\label{s10_52}\\
&& I(-\alpha_\text{c},+\beta_\text{c}, +h_x,+ h_y)=I(+\alpha_\text{c},+\beta_\text{c}, +h_x,- h_y)=
I(-\alpha_\text{c},-\beta_\text{c}, -h_x,+ h_y)=I(+\alpha_\text{c},-\beta_\text{c}, -h_x,- h_y);\; \varphi_0>0,\label{s10_53}\\
&& I(-\alpha_\text{c},-\beta_\text{c}, +h_x,+ h_y)=I(+\alpha_\text{c},-\beta_\text{c}, +h_x,- h_y)=
I(-\alpha_\text{c},+\beta_\text{c}, -h_x,+ h_y)=I(+\alpha_\text{c},+\beta_\text{c}, -h_x,- h_y);\; \varphi_0>0.\label{s10_54}
\end{eqnarray}
\end{subequations}
Finally, to further confirm the obtained functional form for the phase shift, we have set both the cSOC coefficients, i.e., $\alpha_\text{c},\beta_\text{c}$ and the components of Zeeman field, i.e. $h_x,h_y$ to nonzero values. The corresponding CPRs are given in Eqs.~(\ref{I10_set5}). The results illustrate a good agreement with the obtained functional form of phase shift. In particular, the modulus of the phase shift is unchanged in Eqs.~(\ref{s10_51}) and (\ref{s10_54}). The same statement is true for 
Eqs.~(\ref{s10_52}) and (\ref{s10_53}). Furthermore, the modulus of the phase shift in Eq.~(\ref{s10_51}) is larger than that of in Eq.~(\ref{s10_52}). Hence, we arrive at the following functional form for the $\varphi_0$ phase shift in a high-chemical potential regime
\begin{equation}
\varphi_0
\propto \Big( \alpha_\text{c}^2-\Gamma_1\beta_\text{c}^2\Big)h_x\beta_\text{c}+\Big( \alpha_\text{c}^2-\Gamma_2\beta_\text{c}^2\Big)h_y\alpha_\text{c},
\end{equation}
in which the parameters are $\Gamma_2>\Gamma_1$ and $\Gamma_{1,2}>1$. 

\begin{figure}
\centering
{\includegraphics[width=0.49\textwidth]{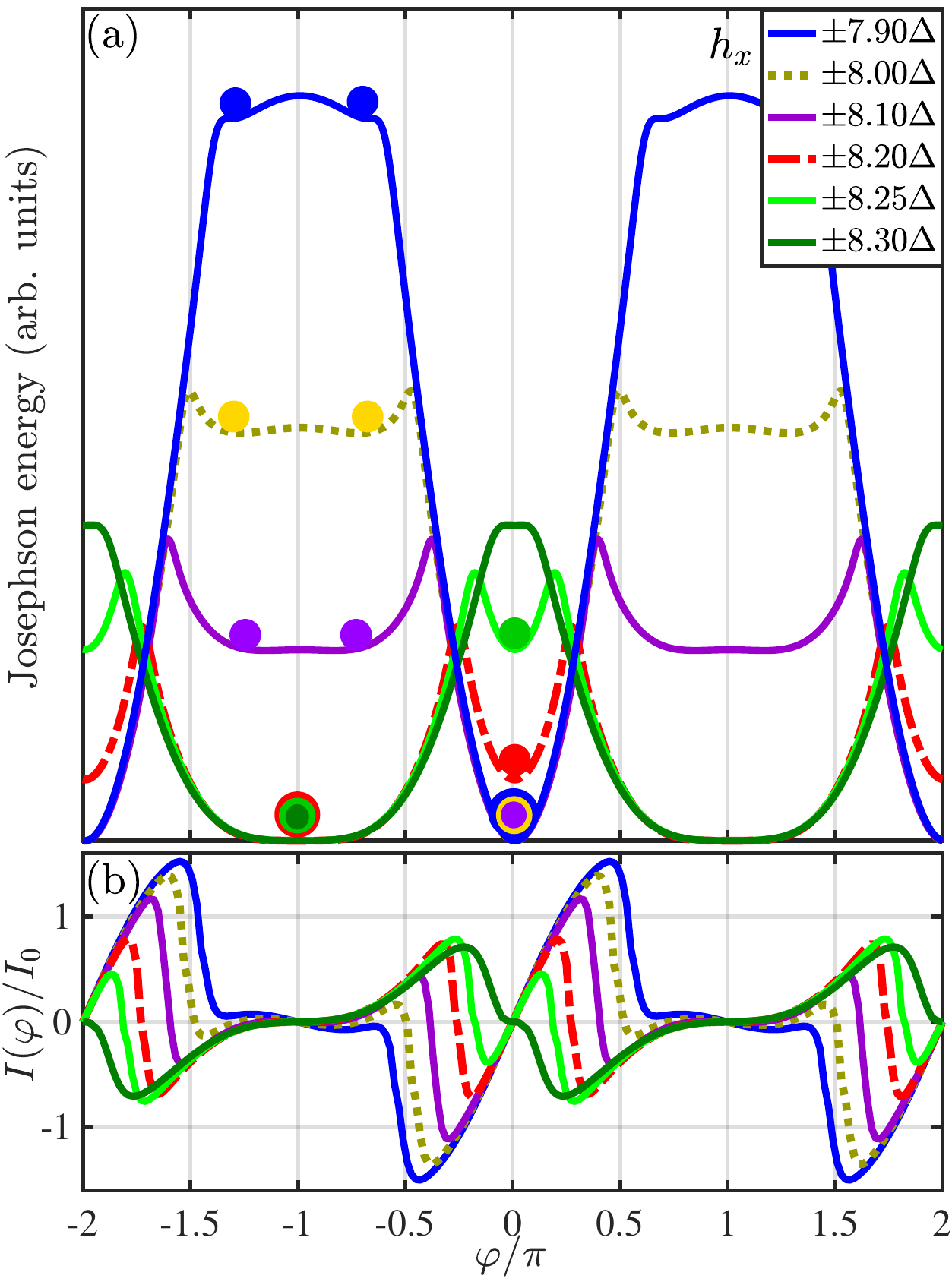}
\includegraphics[width=0.49\textwidth]{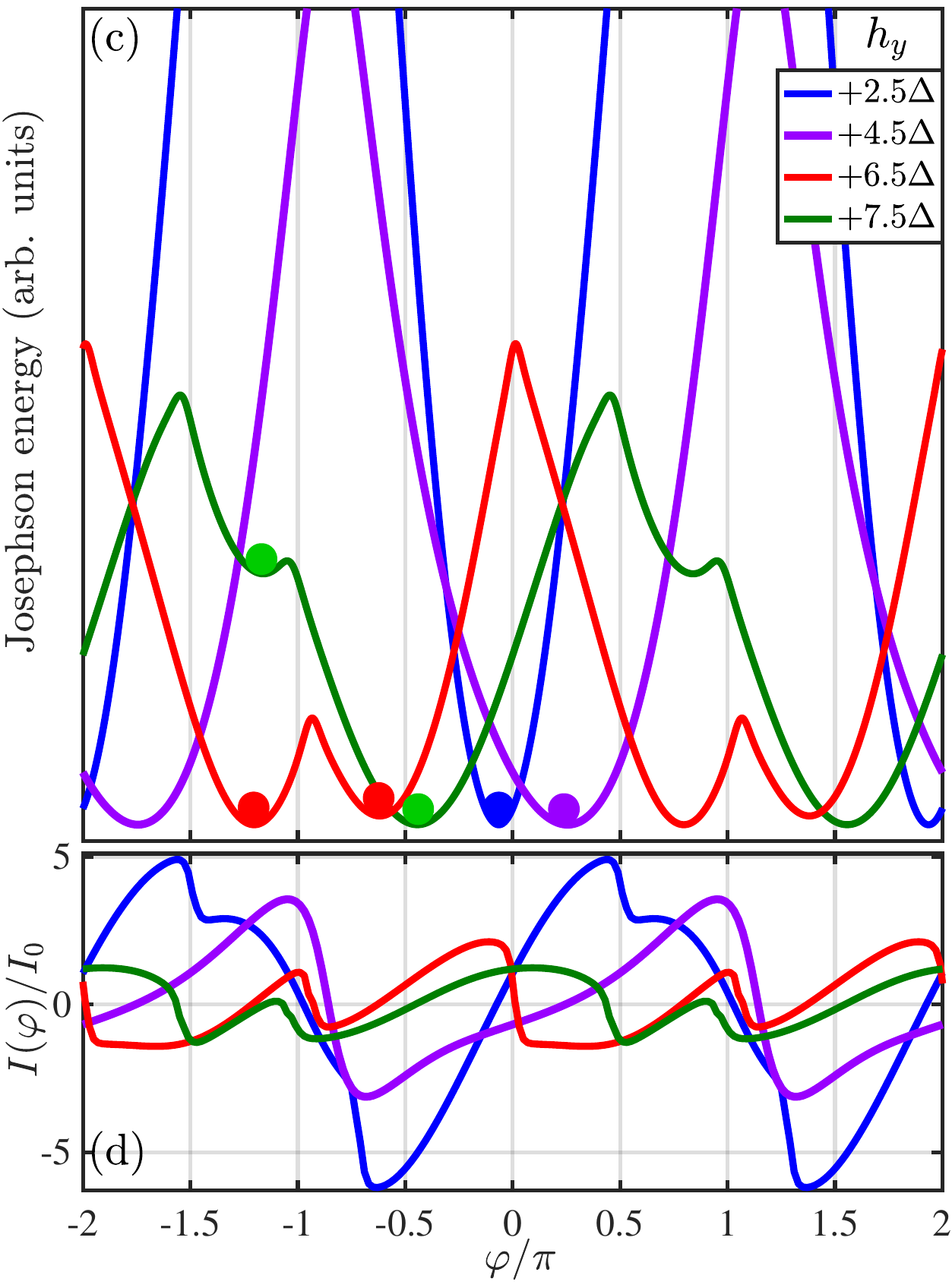}}
\caption{The Josephson energy as a function of superconducting phase difference $\varphi$ (the top row panels) and associated current-phase relations (the bottom row panels). 
In the panels (a), (b) we set the parameter values to $\ac=\pm1$ and $\bc=0$, $\mu=10\Delta$, $h_y=0$,  while in the panels (c), (d) $\ac=1$, $\bc=0$, $\mu=10\Delta$, $h_x=0$. 
Dots/circles denote the global and local minima of the Josephson energy. Various values of the Zeeman field components are considered.}\label{JJE_cpr}
\end{figure}

\begin{figure*}
\centering
\includegraphics[width=\textwidth]{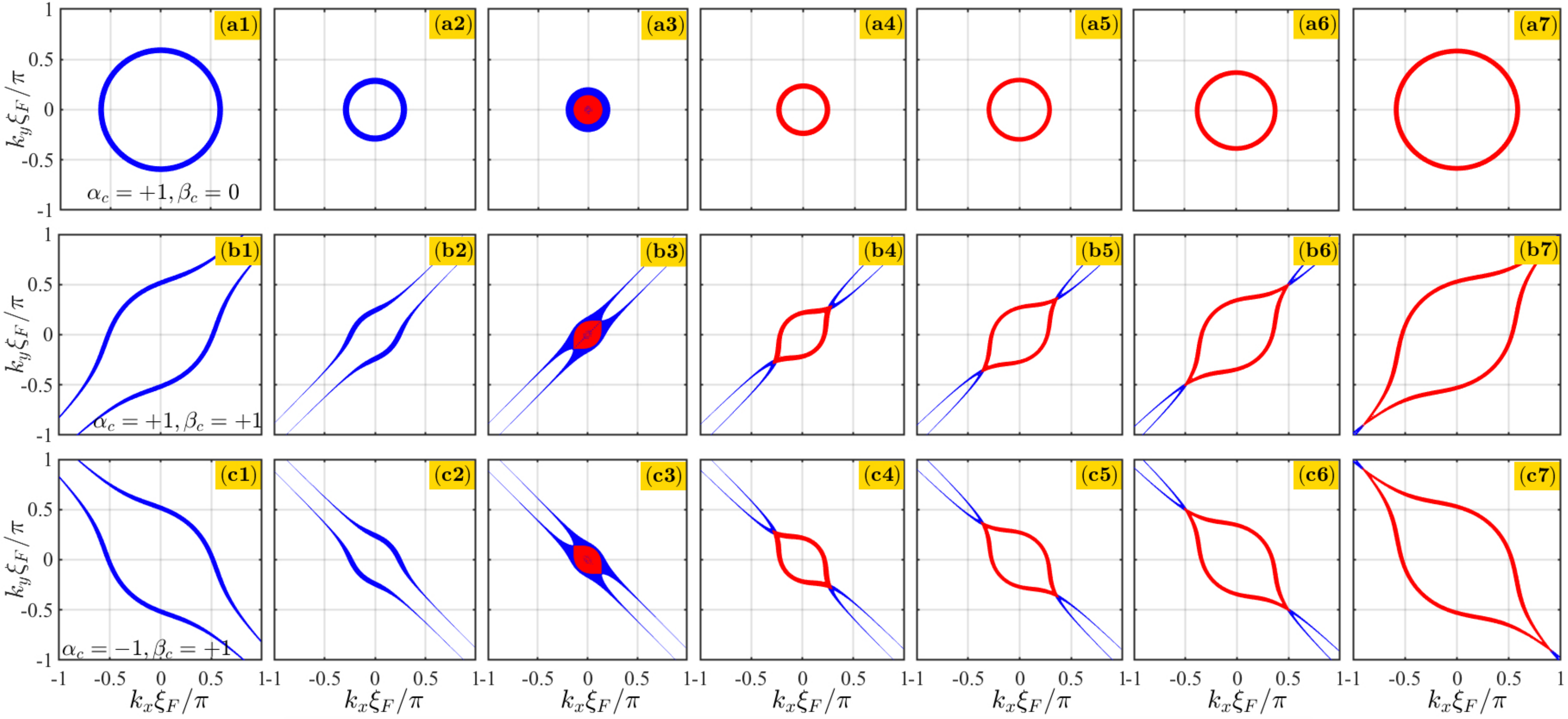}\\
\includegraphics[width=\textwidth]{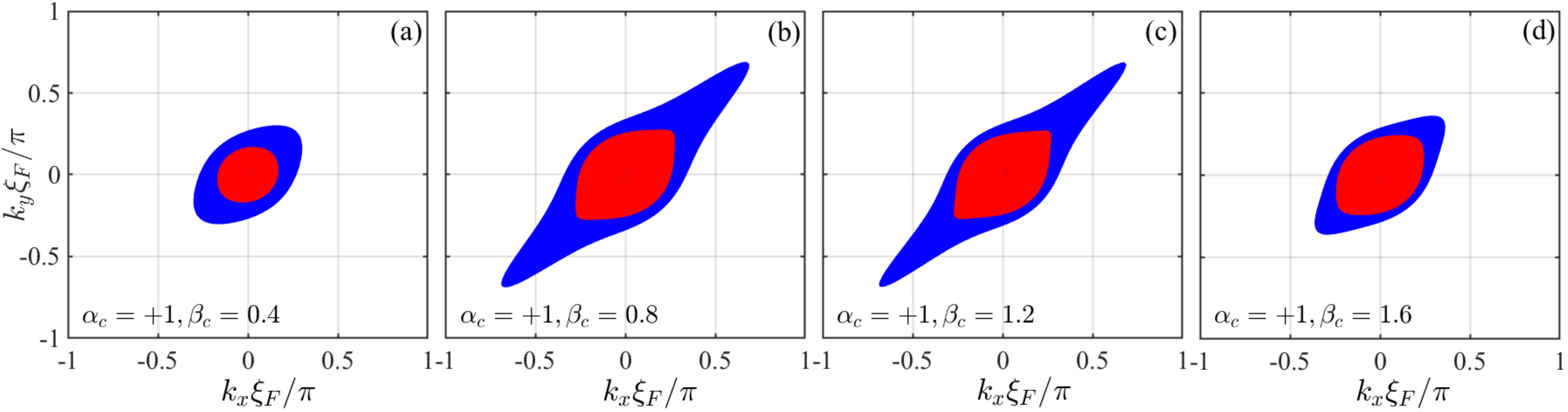}
\caption{Rows 1-3: Isoenergy plots of the band structure, shown for in-plane components of the wave vector and using a Fermi wavelength, 
$\xi_\text{F}=\hbar/\sqrt{2m^*\mu}$. The considered energy interval for the isoenergy surfaces changes column-wise from the left to the right as follows: 
(1) $\Delta E_1$ = -5.5,-4,
(2) $\Delta E_2$ = -0.5, -0.2,
(3) $\Delta E_3$ = -0.1, +0.2,
(4) $\Delta E_4$ = +0.5,+0.9,
(5) $\Delta E_5$ = +1.0,+1.5,
(6) $\Delta E_6$ = +2.0,+2.8,
(7) $\Delta E_7$ = +7.0,+8.8. Rows 4-5: Isoenergy plots of the band structure at the energy interval $\Delta E=-0.3,+0.3$ for different values of the coefficients of cSOC, i.e., $\alpha_\text{c}, \beta_\text{c}$.
}\label{bs1}
\end{figure*}

Using the common description of the two-dimensional electron gas with the Hamiltonian linear in the wave vector and the corresponding Rashba and Dresselhaus coefficients, 
$\alpha_l$, $\beta_l$~\cite{Zutic2004:RMP}, 
from the corresponding results obtained for the $\varphi_0$ state studied in Ref.~\onlinecite{Alidoust2020:PRB1}, one can infer how the interplay between such linear and cubic SOC would alter our 
previous analysis considering only cSOC. Specifically, we find that in the functional form of $\varphi_0$ one needs to replace $\alpha_l$ by $\alpha_l+{\cal A}\alpha_\text{c}$ and $\beta_l$ by $\beta_l+{\cal B}\beta_\text{c}$ in Eq.~(16) of Ref.~\onlinecite{Alidoust2020:PRB1}, where ${\cal A}$ and ${\cal B}$ are two constants dependent on the junction parameters such as the chemical potential. From this analysis we can conclude that $\varphi_0$ state occurs when {\bf h} shifts {\bf p} $\bot$ to ${\bm I}(\varphi)$ and thus alters the SOC, while such occurrence
of $\varphi_0$ does not depend if the SOC has a linear of cubic form.

\section{III. Josephson Energy}
For providing a complete picture of the supercurrent profile partially presented in the main text, Fig.~\ref{JJE_cpr} shows the Josephson energies as a function of $\varphi$ and associated CPRs for a high-chemical potential regime, $\mu=10\Delta$. In Figs.~\ref{JJE_cpr}(a) and \ref{JJE_cpr}(b), cSOC and Zeeman field are not effectively coupled. As can be seen, the Josephson energy possesses global minima either at 
$\varphi=0$ or $\varphi=\pi$ due to the Zeeman field (magnetization) strength $h_x$. Additionally, there are local minima marked by dots. Note that the change in the magnetization orientation  alone is unable to alter the obtained results. Unlike the low-chemical potential regime presented in the main text, we observe that the transition of the global Josephson energy minimum from the $0$ state to the $\pi$ state occurs abruptly 
for $\mu=10\Delta$ despite the presence of higher harmonics than the first harmonic in CPR (see Fig.~\ref{JJE_cpr}(a)). This transition is not accompanied by the intermediate $\varphi_0$ states found for $\mu=\Delta$ in the main text. The corresponding CPRs change sign through the emergence of higher harmonics as seen in Fig.~\ref{JJE_cpr}(b). These higher harmonics, however, are unable to induce a global $\varphi_0 \neq 0$ and only result in local minima in with higher energy values than the ground states in Josephson energy marked by yellow and violet circles in Fig.~\ref{JJE_cpr}(a). 
Nevertheless, Figs.~\ref{JJE_cpr}(c) and \ref{JJE_cpr}(d) illustrate that when cSOC and Zeeman field are effectively coupled, independent $\mu$, the $\varphi_0$ state is controllable through manipulating the magnetization orientation, magnetization strength, and the strength of cSOC parameters.

\section{IV. Band Structure}

Figure~\ref{bs1} shows the band structure associated with the Hamiltonian given in the main text. To visualize how the band structure behaves in momentum space, we have used isoenergy intervals to show the intersections of band structure around specific energies. The blue and red colors represent two different bands. In the top, middle, and bottom row panels the cSOC coefficients are set to '$\alpha_\text{c}=+1,\beta_\text{c}=0 $', '$\alpha_\text{c}=+1,\beta_\text{c}=+1 $', and '$\alpha_\text{c}=-1,\beta_\text{c}=+1$', respectively. As can be seen, in the absence of $\beta_\text{c}$, the band structure has the usual circular shape as a function of $k_{x},k_{y}$, both for the negative and positive values of energy. In the presence of $\beta_\text{c}$, the isoenergy contours take an elliptical shape where the sign of $\alpha_\text{c}$ determines its orientation. Unlike the top-row panels, where the two bands are almost separated in energy (except around $E=0$), the bottom and top bands tend to coexist. 

To show how the increase of $\beta_\text{c}$ reshapes isoenergy contours, we plot the isoenergy band structure for '$\alpha_\text{c}=+1,\beta_\text{c}=0.4 $', '$\alpha_\text{c}=+1,\beta_\text{c}=+0.8 $', '$\alpha_\text{c}=+1,\beta_\text{c}=+1.2 $', and '$\alpha_\text{c}=+1,\beta_\text{c}=+1.6 $' in Figs.~\ref{bs1}(a)-\ref{bs1}(d). To visualize them clearly, we have chosen a wider energy interval, i.e., $\Delta E = -0.3, +0.3$. The elliptic shape starts to form when $\beta_\text{c}\neq 0$, reaches maximum at $\beta_\text{c}=1$, and begins to return back to the original circular shape by further increasing $\beta_\text{c}>1$. 

\section{V. Majorana flat bands} 

\begin{figure*}
\centering
\includegraphics[width=\textwidth]{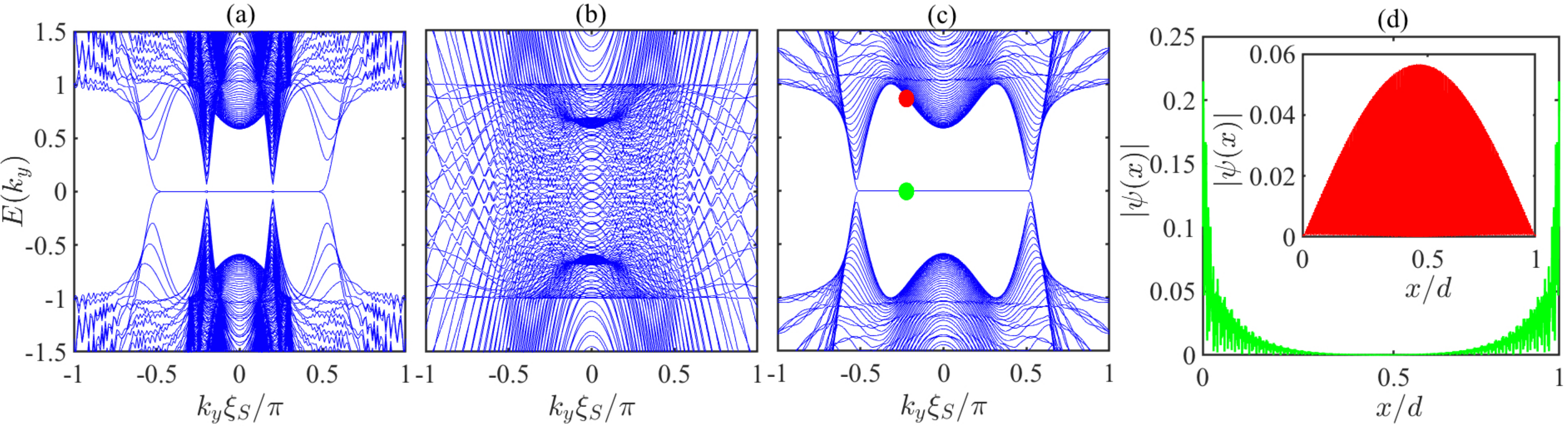}
\caption{\label{MFs}(a)-(c): Band structure of a finite-size system in the $x$ direction as a function of the wave vector $k_y$, using the characteristic length $\xi_\text{S}=\hbar/\sqrt{2m^*\Delta}$. In the panel (a) $\alpha_\text{c}=1.0,\beta_\text{c}=0.0, h_x=2\Delta, h_y=h_z=0$. The same band structure appears when $h_z=\Delta, h_x=h_y=0$. In the panel (b), all parameters are set equal to those of the panel (a) except $h_y=\Delta$ and $h_x=h_z=0$. The band structure in the panel (c) shows results for $\alpha_\text{c}=0.0,\beta_\text{c}=1.0, h_y=2\Delta, h_x=h_z=0$ and $\alpha_\text{c}=0.0,\beta_\text{c}=1.0, h_z=2\Delta, h_x=h_y=0$. When we set $h_x=\Delta$ and $h_y=h_z=0$, a band structure with similar characteristics as the one in the panel (b) appears. In the panel (d), the modulus of wave functions $|\psi(x)|$ as a function of location for two different points (the green and red circles) shown in the panel (c) are plotted. The chemical potential is set to $\mu=\Delta$.
}
\end{figure*}

To demonstrate the existence of zero-energy Majorana flat bands in a system hosting the three ingredients, cSOC, Zeeman field (magnetization), and superconductivity, we plot in Fig.~\ref{MFs}
the band structure of a such system, with a finite size in the $x$ direction, as a function of $k_y$. In this scenario, the y-component of momentum is a good quantum number and remains conserved. In Figs.~\ref{MFs}(a), \ref{MFs}(b), we set $\alpha_\text{c}=1.0$ and $\beta_\text{c}=0.0$,  while $\alpha_\text{c}=0.0$ and $\beta_\text{c}=1.0$ in Figs.~\ref{MFs}(c), \ref{MFs}(d). As can be seen, 
$h_{x,z}\neq 0, h_y=0$ and $h_{y,z}\neq 0, h_x=0$ result in a flat band at zero energy when $\alpha_\text{c}=1.0, \beta_\text{c}=0.0$ and $\alpha_\text{c}=0.0, \beta_\text{c}=1.0$, respectively. Otherwise, the band structure possesses no energy gap. To show the difference between the Majorana flat bands [corresponding to the flat band at $E=0$ in Figs.~\ref{MFs}(a), \ref{MFs}(c)], we plotted 
the modulus of wave function as a function of position in Fig.~\ref{MFs}(d). The wave functions are associated to the two circles displayed in Fig.~\ref{MFs}(c). The wave function of the flat band mode, marked by the green circle, is highly localized at the boundaries $x=0,d$ whereas it propagates throughout the system for the bands marked by the red circle in Fig.~\ref{MFs}(c).

\section{VI. Green function and spatial symmetries of superconducting pair correlations}

The Green function in the Matsubara representation is defined as~\cite{Zagoskin:2014} 
\begin{eqnarray}\label{GF1}
\hat{G}(\omega; \mathbf{r},\mathbf{r}') =\int_0^{\hbar /k_B T} d\tau e^{i\omega \tau} 
\bigg[
 \begin{matrix}
 G(\tau; \mathbf{r},\mathbf{r}')  & F(\tau; \mathbf{r},\mathbf{r}')\\
F^{\dag}(\tau; \mathbf{r},\mathbf{r}') & G^{\dag}(\tau; \mathbf{r},\mathbf{r}')
 \end{matrix}
\bigg],~~~~~~
\end{eqnarray}
where  $\omega =\pi k_BT(2n+1)/\hbar$ is the Matsubara frequency with $T$ as the temperature and $n\in \mathcal{Z}$ and the integration if performed over the imaginary time, $\tau$. 
The normal $G(\tau; \mathbf{r},\mathbf{r}')$ and anomalous $F(\tau; \mathbf{r},\mathbf{r}')$ components of the Green function are defined by 
\begin{subequations}\label{GF_comp}
\begin{align}
&G_{ss'}(\tau; \mathbf{r},\mathbf{r}') =  - \langle T_{\tau} \psi_s(\tau,\mathbf{r}) \psi^{\dag}_{s'}(0,\mathbf{r}')\rangle,\\
&G^{\dag}_{ss'}(\tau; \mathbf{r},\mathbf{r}') =- \sigma^y_{ss_1}\langle T_{\tau} \psi^{\dag}_{s_1}(\tau,\mathbf{r}) \psi_{s_2}(0,\mathbf{r}') \rangle \sigma^y_{s_2s'},\\
&F_{ss'}(\tau; \mathbf{r},\mathbf{r}') = +\langle T_{\tau} \psi_s(\tau,\mathbf{r}) \psi_{s_1}(0,\mathbf{r}') \rangle (-i\sigma^y_{s_1 s'} ), \label{GF_F1}\\
&F^{\dag}_{ss'}(\tau; \mathbf{r},\mathbf{r}') = +i\sigma^y_{ss_1} \langle T_{\tau} \psi^{\dag}_{s_1}(\tau,\mathbf{r}) \psi^{\dag}_{s'}(0,\mathbf{r}')\rangle, \label{GF_F2}
\end{align}
\end{subequations}
where $s, s', s_1$ are spin indices, the summation is implied over $s_1$, $\tau$ is the imaginary time, $\psi_s$ is the field operator, and $T_{\tau}$ denotes time ordering of operators.

For comparison, we have also studied the linear SOC regime, using the common Hamiltonian 
\begin{equation}\label{sol}
H_\text{lSOC}(\mathbf{k}) = \Big(\alpha_l k_y + \beta_l p_x, -\alpha_l p_x - \beta_l p_y, 0\Big)\cdot \boldsymbol{\sigma}, \end{equation}
in which $\alpha_l$ and $\beta_l$ are the Rashba and Dresselhaus coefficients~\cite{Zutic2004:RMP}. Experimentally, $\alpha_l$ is easily controlled externally, for example, by the application of an electric field.

\begin{figure*}
\centering
\includegraphics[width=\textwidth]{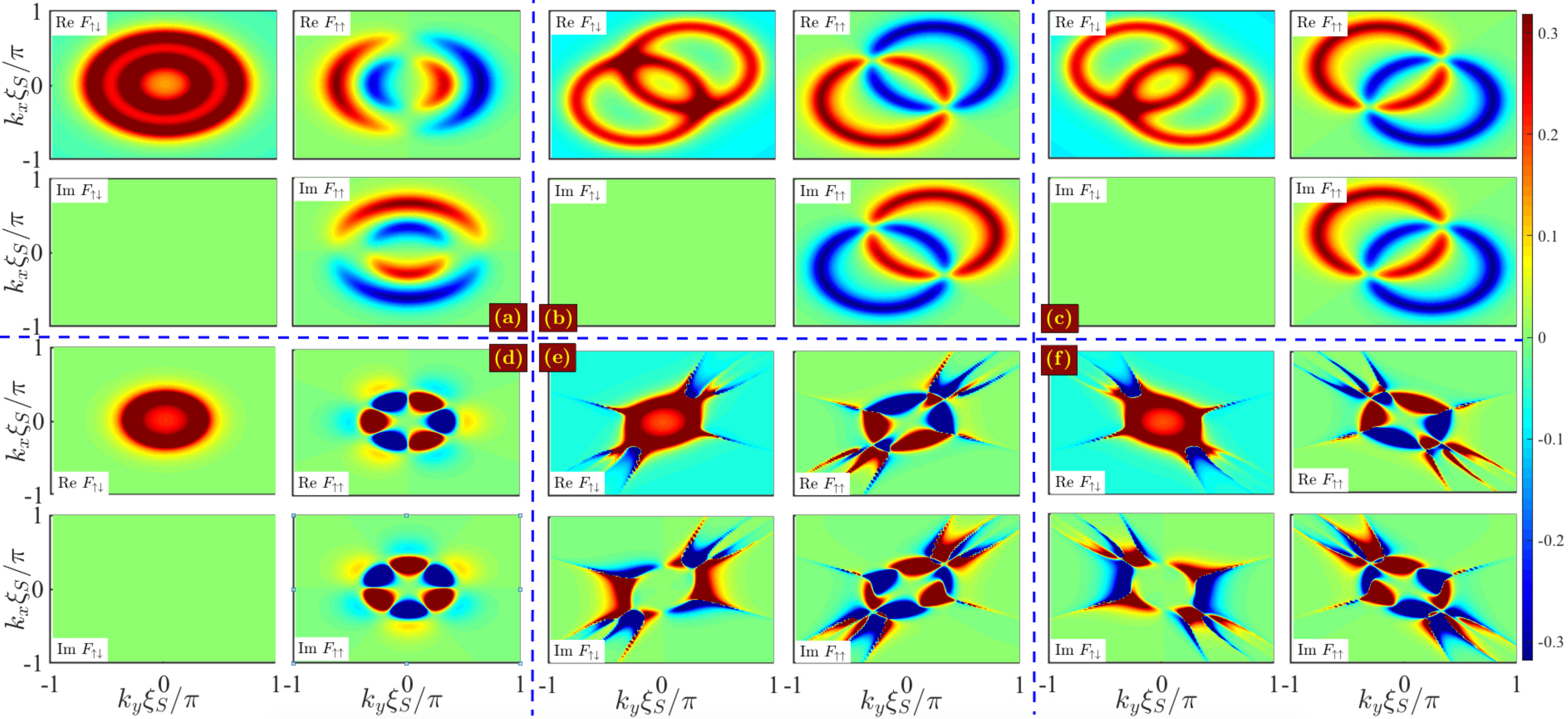}
\caption{\label{GF_fig}Real and imaginary parts of opposite-spin and equal-spin superconducting correlations, $F_{\uparrow\downarrow}$ and $F_{\uparrow\uparrow}$ respectively, as a function of $k_x, k_y$, using the characteristic length $\xi_\text{S}=\hbar/\sqrt{2m^*\Delta}$. In two top rows and two bottom rows linear and cubic SOC is considered, respectively. In the two  left-most columns, two middle columns, and two  right-most columns we set: `$\alpha_{l,\text{c}}=1, \beta_{l,\text{c}}=0$', `$\alpha_{l,\text{c}}=\beta_{l,\text{c}}=1$', and `$\alpha_{l,\text{c}}=- \beta_{l,\text{c}}=1$', respectively. The rest of 
the parameter values are set identical for all the panels.}
\end{figure*}

We briefly discuss momentum-space symmetry profiles of opposite-spin and equal-spin superconducting pair correlations in a system hosting simultaneously Rashba and Dresselhaus SOC. 
To this end, using the Hamiltonian presented in the main text, in the absence of magnetization $\mathbf{h}=0$, we construct the total Green function from 
Eq.~(\ref{GF1}) and derive its components Eq.~(\ref{GF_comp}). We assume that the two-dimensional system is translationally invariant in the $x$ and $y$ directions. This assumption simplifies subsequent analysis and allows for deriving analytical expressions for the Green function in the momentum space. The anomalous components of Green function, Eqs.~(\ref{GF_F1}) and (\ref{GF_F2}), correspond to superconducting pair correlations that exist in the system. Therefore, although we have obtained all the components of the Green function from Eq.~(\ref{GF_comp}) we focus on the anomalous component only. Due to the symmetry among these components, we present results for $F_{\uparrow\uparrow}(\omega; \textbf{k}) $ and $F_{\uparrow\downarrow}(\omega; \textbf{k}) $.
The Green function with $h_x\neq h_y\neq h_z\neq 0$ and $\alpha_{l,\text{c}}\neq\beta_{l,\text{c}}\neq 0$ is highly complicated and lengthy. However, it becomes tractable if we set $\mathbf{h}=0$. The anomalous components of Green function with linear SOC and $\mathbf{h}=0$ read
\begin{subequations}\label{apnx_AGF}
\begin{align}\label{apnx_AGF_a}
&\Omega F_{\uparrow\uparrow}(\omega; \textbf{k})  = 2 \Delta(\beta_l (k_x + i k_y) + \alpha_l (i k_x + k_y))  (\gamma(k_x^2 + k_y^2) - \mu),\\
&\Omega F_{\uparrow\downarrow}(\omega; \textbf{k}) = \Delta (4 \alpha_l \beta_l k_x k_y + \alpha_l^2 (k_x^2 + k_y^2) + 
   \beta_l^2 (k_x^2 + k_y^2) + \Delta^2 + (\gamma(k_x^2 + k_y^2) - \mu)^2 + \omega^2),\\
&\nonumber \Omega = 8 \alpha_l^3 \beta_l k_x k_y (k_x^2 + k_y^2) + \alpha_l^4 (k_x^2 + k_y^2)^2 + 
 \beta_l^4 (k_x^2 + k_y^2)^2 - 
 2 \beta_l^2 (k_x^2 + 
    k_y^2) (-\Delta^2 +  (\gamma(k_x^2 + k_y^2) - \mu)^2 - \omega^2) + 
 8 \alpha_l \beta_l k_x k_y (\beta_l^2 (k_x^2 + k_y^2) + \Delta^2 \nonumber \\& \nonumber - (\gamma(k_x^2 + k_y^2) - \mu)^2 + \omega^2) + (\Delta^2 + \gamma(k_x^2 + k_y^2) - \mu)^2 + \omega^2)^2 + 
 2 \alpha_l^2 (\beta_l^2 (k_x^4 + 10 k_x^2 k_y^2 + k_y^4) - (k_x^2 + 
       k_y^2) (-\Delta^2 + (\gamma(k_x^2 + k_y^2) - \mu)^2 - \omega^2)),
\end{align}
\end{subequations}
where for brevity we use $\gamma=\hbar^2/(2m^*)$  and set $\hbar=1$ in the term involving $\omega$. To further simplify analytical expressions we consider three sets of parameter values: ($i$) $\alpha_l\neq 0$, $\beta_l=0$ and ($ii$, $iii$) $\beta_l=\pm \alpha_l$. The resulting anomalous Green functions within the first parameter set considered (i.e. ($i$) $\alpha_l\neq 0$, $\beta_l=0$), can be expressed by:
\begin{subequations}\label{AGF_set1}
\begin{align}
&\Omega F_{\uparrow\uparrow}(\omega; \textbf{k})  = 2 \Delta\alpha_l (i k_x + k_y) (\gamma(k_x^2 + k_y^2) - \mu),\\
&\Omega F_{\uparrow\downarrow}(\omega; \textbf{k}) = \Delta (\alpha_l^2 (k_x^2 + k_y^2) + \Delta^2 + (\gamma(k_x^2 + k_y^2) - \mu)^2 + \omega^2),\\
&\Omega = \alpha_l^4 (k_x^2 + k_y^2)^2 - 
 2 \alpha_l^2 (k_x^2 + 
    k_y^2) (-\Delta^2 + (\gamma(k_x^2 + k_y^2) - \mu)^2 - \omega^2)+ (\Delta^2 + (\gamma(k_x^2 + k_y^2) - \mu)^2 + \omega^2)^2. \nonumber
\end{align}
\end{subequations}
Additionally, or ($ii$, $iii$) $\beta_l=\pm \alpha_l$ the Green function reads
\begin{subequations}\label{AGF_set2}
\begin{align}
&\Omega F_{\uparrow\uparrow}(\omega; \textbf{k})  = \pm 2\alpha_l \Delta (1 \pm  i)  (k_x \pm k_y)  (\gamma(k_x^2 + k_y^2) - \mu),\\
&\Omega F_{\uparrow\downarrow}(\omega; \textbf{k}) = \Delta (2 \alpha_l^2 (k_x \pm k_y)^2 + \Delta^2 + (\gamma(k_x^2 + k_y^2) - \mu)^2 + \omega^2),\\
&\Omega = 4 \alpha_l^4 (k_x \pm k_y)^4 - 
 4 \alpha_l^2 (k_x \pm 
    k_y)^2 (-\Delta^2 + (\gamma(k_x^2 + k_y^2) - \mu)^2 - \omega^2) + (\Delta^2 + (\gamma(k_x^2 + k_y^2) - \mu)^2 + \omega^2)^2.
\end{align}
\end{subequations}
The same study with cSOC, given in the main text, results in
\begin{subequations}\label{GFc}
\begin{align}
&\Omega F_{\uparrow\uparrow} = 2 \Delta (k_x - i k_y)^2 (-\alpha_\text{c}(i k_x + k_y) + \beta_\text{c}(k_x + 
   i k_y) (\gamma(k_x^2 + k_y^2)-\mu),\\
 &\Omega F_{\uparrow\downarrow} = \Delta ((k_x^2 + k_y^2)^2 (-4 \alpha_\text{c} \beta_\text{c} k_x k_y + 
      (\alpha_\text{c}^2 + \beta_\text{c}^2)(k_x^2  + 
      k_y^2 ) + \gamma^2) + \Delta^2 \
- 2 \gamma \mu (k_x^2 + k_y^2)  + \mu^2 + \omega^2),\\
&\Omega = -(-i ak_x + ak_y + bk_x - 
     i bk_y) (-(i ak_x + ak_y + bk_x - i bk_y) \Delta^2 + (i ak_x +
        ak_y + bk_x + 
       i bk_y) (-(-i ak_x + ak_y + bk_x - i bk_y) \nonumber \\ & (i ak_x + ak_y + bk_x - 
          i bk_y) + (-\gamma k^2 + \mu + 
         i \omega)^2)) +  (\Delta^2 (\gamma k^2 - \mu - 
       i \omega) + (-(-i ak_x + ak_y + bk_x - i bk_y) (i ak_x + ak_y + 
          bk_x - i bk_y) +  \nonumber \\ &(-\gamma k^2 + \mu + 
         i \omega)^2) (\gamma k^2 - \mu + 
       i \omega)) (\gamma k^2 - \mu + 
    i \omega) + \Delta^2 ((ak_y + bk_x)^2 + (ak_x + 
      bk_y)^2 + \Delta^2 + (-\gamma k^2  + \mu)^2 +
\omega^2),
\end{align}
\end{subequations}
where, to further simplify the expressions, we have defined new variables $ak_x = \alpha_\text{c}(k_x^3 - 3 k_xk_y^2)$, $ak_y = \alpha_\text{c}(k_y^3 - 3 k_yk_x^2)$, $bk_x = \beta_\text{c}(k_x^3 + k_xk_y^2)$, $bk_y =  \beta_\text{c}(k_y^3 + k_yk_x^2)$, $k^2 = k_x^2 + k_y^2$. As can be seen, even by introducing these simplifying variables, the denominator of the Green function, $\Omega$, is highly complicated. Therefore, in what follows, we present only the numerator of the anomalous Green function. For the case ($i$) $\alpha_c\neq 0$, $\beta_c=0$ we find 
\begin{subequations}\label{GFc2}
\begin{align}
   &\Omega F_{\uparrow\uparrow} =-2 \alpha_\text{c}\Delta (k_x - i k_y)^2 (i k_x + k_y) (\gamma(k_x^2 + k_y^2)-\mu),\\
   &\Omega F_{\uparrow\downarrow} =\Delta ((k_x^2 + k_y^2)^2 (\alpha_\text{c}^2(k_x^2  + 
      k_y^2 ) + \gamma^2) + \Delta^2 \
- 2 \gamma \mu (k_x^2 + k_y^2)  + \mu^2 + \omega^2),
\end{align}
\end{subequations}
whereas for the cases ($ii,iii$) $\beta_\text{c}=\pm\alpha_\text{c}$ the anomalous components of Green function reduce to
\begin{subequations}\label{GFc3}
\begin{align}
   &\Omega F_{\uparrow\uparrow} =\pm 2\alpha_\text{c} \Delta(1 \mp i) (k_x \mp k_y) (k_x - 
   i k_y)^2  ((k_x^2 + k_y^2) \gamma- \mu),\\
   &\Omega F_{\uparrow\downarrow} =\Delta ((k_x^2 + 
      k_y^2)^2 (2 \alpha_\text{c}^2(k_x \mp 
         k_y)^2  + \gamma^2) + \Delta^2 - 
   2 \gamma \mu (k_x^2 + k_y^2)  + \mu^2 + \omega^2).
\end{align}
\end{subequations}

As can be seen from the above relations, the obtained Green functions are too complicated to be analyzed analytically. Therefore, we evaluate these expressions numerically to provide a detailed overview of these superconducting correlations. Figure \ref{GF_fig} illustrates the profile of the opposite-spin ($F_{\uparrow\downarrow}$) and equal-spin ($F_{\uparrow\uparrow}$) superconducting correlations in momentum-space. We have presented both the real and imaginary parts of Green functions. The energies are normalized by the superconducting gap $\Delta$. The chemical potential is fixed at $\mu=2\Delta$ and Matsubara frequency at zeroth mode, i.e, $n=0$, as representative values. In Figs.~\ref{GF_fig}(a)-\ref{GF_fig}(c) we consider the linear SOC, while in Figs.~\ref{GF_fig}(d)-\ref{GF_fig}(f) the cSOC is studied. To facilitate comparisons, we set $\alpha_{l,\text{c}}=1, \beta_{l,\text{c}}=0$ in Figs.~\ref{GF_fig}(a) and \ref{GF_fig}(d), $\alpha_{l,\text{c}}=\beta_{l,\text{c}}=1$ in Figs.~\ref{GF_fig}(b) and \ref{GF_fig}(e), and $\alpha_{l,\text{c}}=- \beta_{l,\text{c}}=1$ in Figs.~\ref{GF_fig}(c) and \ref{GF_fig}(f). By setting the above specific choices, we simplify further the anomalous Green function as given by Eqs.~(\ref{AGF_set1})-(\ref{AGF_set2}) and (\ref{GFc2})-(\ref{GFc3}). It can be seen that the opposite-spin component has an effective even-parity symmetry ($s$-wave class), while the equal-spin component possesses effective odd-parity symmetry [$p$-wave (linear SOC) and $f$-wave (cSOC) classes]. Employing the full expressions of the anomalous Green function, Eq.~(\ref{apnx_AGF}), we have numerically confirmed these effective symmetries in Figs.~\ref{GF_fig}(a)-\ref{GF_fig}(c). The negative and positive values are shown by deep blue and red, respectively. Hence, it is apparent that the effective $s$-wave symmetry class experiences no sign change by $k\rightarrow-k$, while the effective $p$-wave symmetry class shows a sign change upon $k\rightarrow-k$. The change of $\alpha_l=+\beta_l$ to $\alpha_l=-\beta_l$ rotates the orientation of ring-like symmetries in Figs.~\ref{GF_fig}(b) and \ref{GF_fig}(c). The model employed for the linear SOC, Eq.~(\ref{sol}), results in $k_x=\pm k_y$ symmetry lines. In Figs.~\ref{GF_fig}(d)-\ref{GF_fig}(f), we have evaluated the anomalous Green functions Eqs.~(\ref{GFc}). As can be seen in Figs.~\ref{GF_fig}(d), when $\alpha_\text{c}\neq 0, \beta_\text{c}=0$,  the opposite-spin Green function, $F_{\uparrow\downarrow}$, shows the usual $s$-wave symmetry as a function of momenta whereas the equal-spin Green function, $F_{\uparrow\uparrow}$, exhibits an $f$-wave symmetry. Such an $f$-wave symmetry has appeared in bilayer graphene and black phosphorus where these symmetries along with $d$-wave and extended $s$-wave are found to be interchangeable upon mechanical displacement of graphene layers or the application of an external strain~\cite{Alidoust2020:PRB2,Alidoust2018:PRB2,Alidoust2018:PRB3}. Introducing nonzero $\beta_\text{c}\neq 0$ in Figs.~\ref{GF_fig}(e) and \ref{GF_fig}(f) with $\alpha_\text{c}=\pm\beta_\text{c}=1$, respectively, the symmetric pictures in the absence of $\beta_\text{c}$ is now distorted. The $s$-wave symmetry class of $F_{\uparrow\downarrow}$ and the $f$-wave symmetry class of $F_{\uparrow\downarrow}$ now gain specific directions. Similar to the linear SOC, the symmetries in momentum space are oriented around $k_x=\pm k_y$ lines.
Therefore, considering our results above, this property provides an external knob for having control over the orientation of the parity symmetry of superconducting correlations in a Rashba-Dresselhaus 
SOC platform. To 
determine the predicted symmetries of superconducting pair correlations and their spatial orientations, one can employ high-resolution angular point-contact spectroscopy experiments or spatially-resolved Meissner-effect experiments~\cite{Bae2019:RSI,Zutic1997:PRB,Halterman2001:PRB}. 
Since linear and cubic SOC are characterized by different superconducting correlations and different functional forms of $\varphi_0$ (recall Section II),
this is useful for the future efforts to experimentally decouple SOC contributions in the superconducting response.   

\begin{figure*}
\centering
\includegraphics[width=0.8\textwidth]{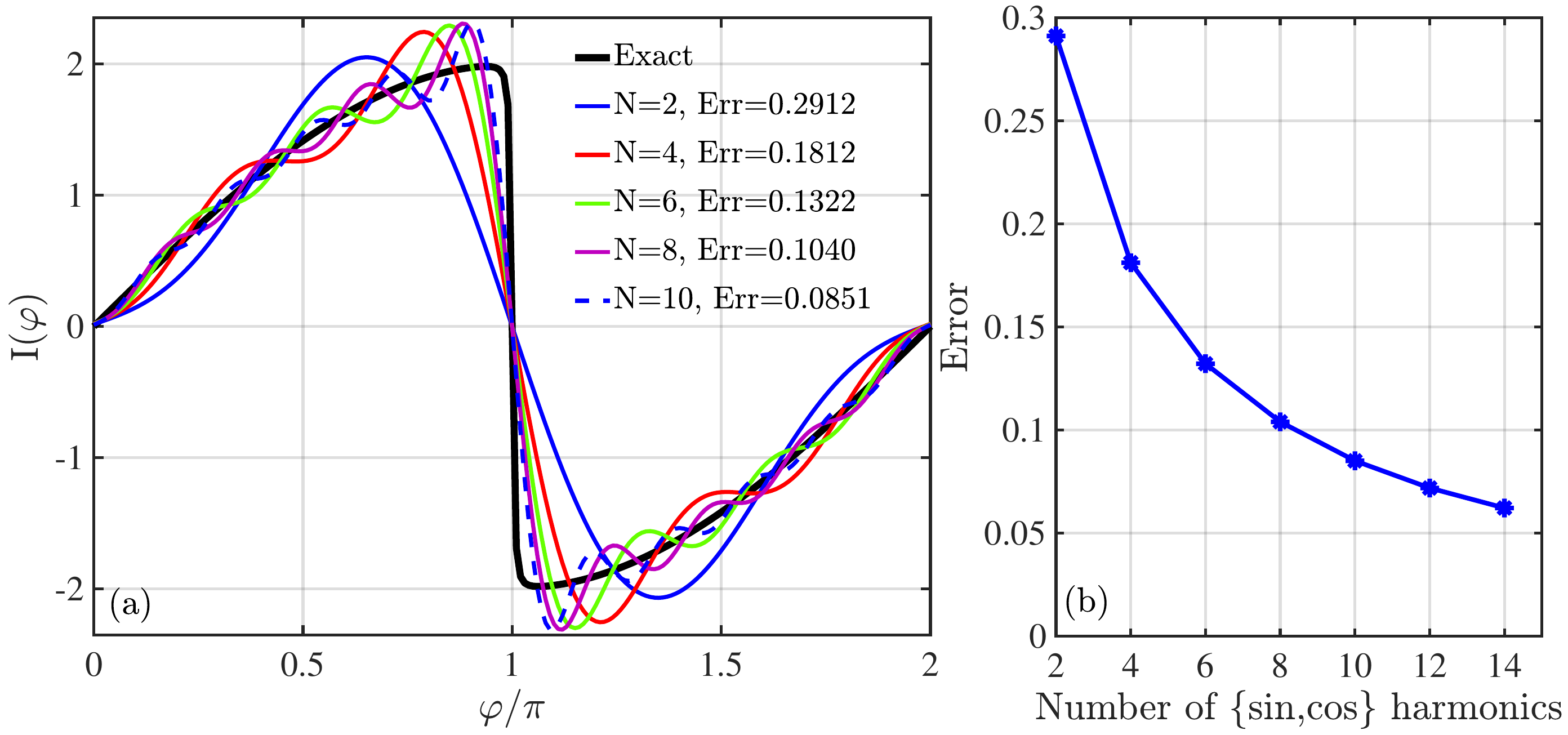}
\caption{\label{cpr_fit}(a) A sample supercurrent profile as a function of superconducting phase difference $\varphi$. The solid black curve denotes
the original current-phase relation and the remaining curves show fitted functions by a $\{\sin,\cos\}$ complete basis set with first N harmonics. (b) The relative error of the fitted curve against the number of harmonics used in the fitting process. 
}
\end{figure*}

\section{VII. Anharmonic current-phase relation and \{$\sin,\cos$\} expansion}
To illustrate how well is the complete set of \{$\sin,\cos$\} basis functions in recovering the CPR, we have shown such study in Fig.~\ref{cpr_fit}. A sample CPR is [marked by the solid black curve in Fig.~\ref{cpr_fit}(a)] is taken as the original function. We then make use of the following expansion to find a best fit to the solid black curve,
\begin{equation}\label{I_sin_cos}
I(\varphi)\approx\sum_{n=1}^N\left[ I_{sn}\sin(n\varphi+\varphi_{sn})+I_{cn}\cos(n\varphi+\varphi_{cn})\right].
\end{equation} 
Here, $I_{sn},I_{cn},\varphi_{sn},\varphi_{cn}$ are considered as fitting parameters and are determined through a Monte-Carlo search sampling. The relative error is defined as the absolute difference between the original and fitted functions divided by the absolute value of the original function. The variable `$n$' runs from $1$ to $N$ and includes up to N first harmonics. 
In Fig.~\ref{cpr_fit}(a), we have shown the results of this fitting process for $N=2,4,6,8,10$ with associated relative errors. As can be seen, even with N=10 (10 $\sin$ functions and 10 $\cos$ functions), the expansion (\ref{I_sin_cos}) is unable to generate a reasonable fitting function to the original CPR. To illustrate how the relative error evolves with increasing the number of included harmonics, Fig.~\ref{cpr_fit}(b) displays the relative error function as a function of N. The relative error follows an exponential decay against N. Therefore, to generate an expansion using \{$\sin,\cos$\} functions (\ref{I_sin_cos}) 
with a relative error less than $0.05$, one needs to include more than N=20 harmonics.

\twocolumngrid

\end{document}